\documentclass{emulateapj}

\usepackage{graphicx}
\usepackage{amssymb}
\usepackage{longtable}
\usepackage{epsf}
\usepackage{apjfonts}
\usepackage{rotating}

\slugcomment{Accepted in the Astrophysical Journal}

\shorttitle{Massive galaxies at $3.0 \leq z<4.0$ from the NMBS}
\shortauthors{Marchesini et al.}

\begin{document}

\title{The Most Massive Galaxies at $3.0 \leq z<4.0$ in the NEWFIRM Medium-Band Survey: Properties and Improved Constraints on the Stellar Mass Function}

\author{
Danilo~Marchesini\altaffilmark{1,9},
Katherine E. Whitaker\altaffilmark{2,9},
Gabriel Brammer\altaffilmark{2,9},
Pieter~G.~van Dokkum\altaffilmark{2,9},
Ivo Labb\'e\altaffilmark{3,9},
Adam Muzzin\altaffilmark{2,9},
Ryan~F.~Quadri\altaffilmark{4,9},
Mariska Kriek\altaffilmark{5,9},
Kyoung-Soo Lee\altaffilmark{6,9},
Gregory Rudnick\altaffilmark{7,9},
Marijn Franx\altaffilmark{4},
Garth D. Illingworth\altaffilmark{8} 
and David Wake\altaffilmark{2}}

\altaffiltext{1}{Department of Physics and Astronomy, Tufts University, Medford, MA 06520, USA}
\altaffiltext{2}{Department of Astronomy, Yale University, New Haven, CT 06520-8101, USA}
\altaffiltext{3}{Carnegie Observatories, Pasadena, CA 91101, USA }
\altaffiltext{4}{Sterrewacht Leiden, Leiden University, NL-2300 RA Leiden, The Netherlands}
\altaffiltext{5}{Department of Astrophysical Sciences, Princeton University, Princeton, NJ 08544, USA}
\altaffiltext{6}{Yale Center for Astronomy and Astrophysics, Departments of Physics and Astronomy, Yale University, New Haven, CT 06520, USA}
\altaffiltext{7}{Department of Physics and Astronomy, University of Kansas, Lawrence, KS 66045, USA}
\altaffiltext{8}{UCO/Lick Observatory, University of California, Santa Cruz, CA 95064, USA}
\altaffiltext{9}{Visiting Astronomer, Kitt Peak National Observatory, 
National Optical Astronomy Observatory, which is operated by the 
Association of Universities for Research in Astronomy (AURA) under 
cooperative agreement with the National Science Foundation.}

\begin{abstract}
We use the optical to mid-infrared coverage of the NEWFIRM Medium-Band 
Survey (NMBS) to characterize, for the first time, the properties of 
a mass-complete sample of 14 galaxies at $3.0 \leq z < 4.0$ with 
$M_{\rm star}>2.5\times 10^{11}$~M$_{\sun}$, and to derive significantly more 
accurate measurements of the high-mass end of the stellar mass function 
(SMF) of galaxies at $3.0 \leq z < 4.0$. The accurate photometric 
redshifts and well-sampled SEDs provided by the NMBS combined with the 
large surveyed area result in significantly reduced contributions from 
photometric redshift errors and cosmic variance to the total error budget 
of the SMF. The typical very massive galaxy 
at $3.0 \leq z < 4.0$ is red and faint in the observer's optical, with a 
median $r$-band magnitude of $\langle r_{\rm tot}\rangle=26.1$, and median 
rest-frame $U-V$ colors of $\langle U-V\rangle=1.6$. About 60\% of the 
mass-complete sample have optical colors satisfying either the $U$- or 
the $B$-dropout color criteria, although $\sim$50\% of these galaxies 
have $r>25.5$. We find that $\sim$30\% of the sample has SFRs from SED 
modeling consistent with zero, although SFRs of up to 
$\sim1-18$~M$_{\sun}$~yr$^{-1}$ are also allowed within 1~$\sigma$. However, 
$>80$\% of the sample is detected at 24~$\mu$m, resulting in total 
infrared luminosities in the range (0.5-4.0)$\times$10$^{13}$~L$_{\sun}$. 
This implies the presence of either dust-enshrouded starburst activity 
(with SFRs of 600-4300~M$_{\sun}$~yr$^{-1}$) and/or highly-obscured active 
galactic nuclei (AGN). The contribution of galaxies with 
$M_{\rm star}>2.5\times 10^{11}$~M$_{\sun}$ to the total stellar mass budget at 
$3.0 \leq z < 4.0$ is $\sim8^{+13}_{-3}$\%. Compared to recent estimates of 
the stellar mass density in galaxies with 
$M_{\rm star}\approx10^{9}-10^{11}$~M$_{\sun}$ at $z\sim5$ and $z\sim6$, we find 
an evolution by a factor of 2-7 and 3-22 from $z\sim5$ and $z\sim6$, 
respectively, to $z=3.5$. The previously found disagreement at 
the high-mass end between observed and model-predicted SMFs is now 
significant at the 3~$\sigma$ level when only random uncertainties are 
considered. However, systematic uncertainties dominate the total error 
budget, with errors up to a factor of $\sim8$ in the densities at the 
high-mass end, bringing the observed SMF in marginal agreement with the 
predicted SMF. Additional systematic uncertainties on the high-mass end 
could be potentially introduced by either 1) the intense star-formation 
and/or the very common AGN activities as inferred from the MIPS~24~$\mu$m 
detections, and/or 2) contamination by a significant population of 
massive, old, and dusty galaxies at $z\sim2.6$.  
\end{abstract}



\section{INTRODUCTION}\label{sec-in}

Understanding the formation mechanisms and evolution with cosmic time 
of galaxies is one of the major goals of observational cosmology. An 
effective approach to understand the physical processes governing the 
assembly of galaxies (and their relative importance as a function of 
cosmic time) is to directly measure the growth of the stellar mass in 
galaxies. Galaxies can build their stellar mass both from in-situ star 
formation and/or merger events. 
The mean space density of galaxies per unit stellar mass, or stellar 
mass function (SMF), is one of the most fundamental of all cosmological 
observables. The shape of the SMF retains the imprint of galaxy 
formation and evolution processes. Therefore, the SMF and its evolution 
with cosmic time represent a powerful tool to constrain the physical 
mechanisms regulating the assembly and the evolution of galaxies. 

In the past decade, significant observational progress has been made 
in the measurement of the SMF of galaxies and its evolution with 
redshift. Using photometric redshifts derived from multi-waveband imaging 
surveys, measurements of the SMF of galaxies are now routinely performed 
out to $z\sim5$ (e.g., \citealt{dickinson03}; \citealt{conselice05}; 
\citealt{drory05}; \citealt{fontana06}; \citealt{elsner08}; \citealt{perez08}; 
\citealt{kajisawa09}; \citealt{marchesini09}). The general consensus is 
that at $z>1$ the stellar mass assembly proceeds much more quickly than 
at lower redshifts. In particular, very recent measurements at $z<4$ show 
a dramatic evolution of the SMF of galaxies with redshift as well as 
evidence of mass-dependent evolution of the SMF, with the low-mass end 
evolving more rapidly than the high-mass end (i.e., \citealt{perez08}; 
\citealt{marchesini09}).

Measurements of the SMF have been extended also to even larger redshifts 
($z\sim$4-7; e.g., \citealt{mclure09}; \citealt{stark09}), providing 
estimates of the stellar mass content of the universe when it was only 
$\sim$800~Myr old. However, most of these studies only target Lyman break 
galaxies (LBGs), hence resulting in a potentially biased view of the universe 
against massive and evolved galaxies at such high redshifts. Whereas the 
discovery of a population of very massive and evolved galaxies at 
$z\ga 5$ has now been claimed by several groups (e.g., 
\citealt{yan06}; \citealt{rodighero07}; \citealt{wiklind08}; 
\citealt{mancini09}), convincing evidence for the existence of galaxies 
with $M_{\rm star}>3\times10^{11}$~M$_{\sun}$ at $z>4$ is still missing (e.g., 
\citealt{dunlop07}). 

Closely related to this issue is the very intriguing finding that the 
number density of the most massive galaxies 
($M_{\rm star}> 3\times 10^{11}$~M$_{\sun}$) seems to evolve very little from 
$z\sim4$ to $z\sim1.5$ \citep{marchesini09}, suggesting that the most 
massive galaxies in the universe were mostly in place already at 
$z\sim3.5$, and implying potentially severe disagreements with the 
predictions from the latest generations of semi-analytic models of galaxy 
formation.

However, uncertainties on the observed SMF are still large, especially at 
the high-mass end and at high redshifts \citep{marchesini09}. At 
$z\lesssim3$, the SMF error budget is now almost entirely dominated by 
systematic uncertainties caused by the different SED-modeling assumptions 
adopted to derive stellar masses (e.g., stellar population synthesis 
models or initial mass function (IMF)). Progress in reducing the impact 
of these systematic uncertainties necessarily requires better calibrations 
of the stellar mass estimates (e.g., through measurements of the dynamical 
masses from studies of their kinematics). At $3<z<4$, instead, the 
contributions of photometric redshift errors, small-number statistics, and 
sample variance (due to the relatively small probed volumes) are still 
significant, and dominate the total error budget at the high-mass end of 
the SMF.

In this paper we take advantage of the high-quality data from the ultra-violet 
to the mid-infrared (MIR) available through the NEWFIRM Medium-Band Survey 
(NMBS; \citealt{vandokkum09b}) to derive more accurate measurements of the 
high-mass end of the SMF of galaxies at $3.0 \leq z < 4.0$ by significantly 
reducing the impact of random uncertainties and to characterize, for the 
first time, the observed and rest-frame properties of a mass-selected 
sample of galaxies at $3.0 \leq z < 4.0$. These results are made possible 
by the combination of accurate photometric redshifts and well-sampled 
spectral energy distributions (SEDs) of $K$-selected galaxies at $z>1.5$ 
delivered by the medium-band near-infrared (NIR) filters of the NMBS, as 
well as by its large surveyed area ($\sim$0.5~square degree).

This paper is structured as follows. In \S~\ref{sec-ss} we present the 
mass-selected ($M_{\rm star}>2.5\times10^{11}$~M$_{\sun}$) sample used to measure 
the SMF of galaxies at $3.0 \leq z < 4.0$; in \S~\ref{sec-prop} we present 
the observed and rest-frame properties of the galaxies in the mass-selected 
sample. The stellar mass function and densities of massive galaxies at 
$3.0 \leq z < 4.0$ are presented in \S~\ref{sec-mf}, while in 
\S~\ref{sec-dustold} the systematic effects caused by systematic uncertainties 
in the photometric redshift estimates are quantified. The results are 
summarized in \S~\ref{sec-disc}. We assume $\Omega_{\rm M}=0.3$, 
$\Omega_{\rm \Lambda}=0.7$, and $H_{\rm 0}=70$~km~s$^{-1}$~Mpc$^{-1}$ throughout 
the paper. All magnitudes are on the AB system.


\section{SAMPLE SELECTION}\label{sec-ss}


\subsection{The NEWFIRM Medium Band Survey}

The sample is selected from the NMBS, a moderately wide, moderately 
deep near-infrared imaging survey \citep{vandokkum09b}. The survey 
uses the NEWFIRM camera on the Kitt Peak 4m telescope. The camera 
images a 28$^{\prime} \times$28$^{\prime}$ field with four arrays with a 
native pixel size of 0$^{\prime \prime}$.4. We developed a custom NIR 
filter system for NEWFIRM, comprised of five medium bandwidth filters in 
the wavelength range 1~$\mu$m--1.7~$\mu$m. As shown in \citet{vandokkum09b}, 
these filters pinpoint the Balmer/4000~\AA~ breaks of galaxies at 
$1.5<z\lesssim3.5$, providing accurate photometric redshifts and improved 
measurements of the stellar population parameters. The survey targets two 
28$^{\prime} \times$28$^{\prime}$ fields: a subsection of the COSMOS field 
\citep{scoville07}, and a field containing part of the AEGIS strip 
\citep{davis07}. Field positions and other information are given in 
\citet{vandokkum09b}. Both fields have excellent supporting data, including 
extremely deep optical $ugriz$ data from the CFHT Legacy 
Survey\footnote{\url{http://www.cfht.hawaii.edu/Science/CFHTLS/}} and deep 
Spitzer IRAC and MIPS imaging (\citealt{barmby08}; \citealt{sanders07}). 
{\it Spitzer}-IRAC and MIPS photometry have been added following the procedure 
described in \citet{wuyts07}, which uses a source-fitting algorithm 
developed by I. Labb\'e et al. (2010, in preparation) especially suited 
for heavily confused images for which a higher resolution prior (in this 
case the $K$-band image) is available.\footnote{The IRAC fluxes measured 
in this work have been compared with the publicly available IRAC 
photometry over COSMOS 
(\url{http://irsa.ipac.caltech.edu/data/COSMOS/tables/scosmos/}; 
\citealt{ilbert09}) and AEGIS (\url{http://www.cfa.harvard.edu/irac/egs/}; 
\citealt{barmby08}). The agreement is excellent, with systematic differences 
of $\sim$2\%.} 
Reduced CFHT mosaics were kindly provided to us by the CARS team 
(\citealt{erben09}; \citealt{hildebrandt09}). Additionally, in the COSMOS 
field, we include deep Subaru images with the $B_{\rm J}V_{\rm J}r^{+}i^{+}z^{+}$ 
broad-band filters \citep{capak07}, Subaru images with 12 intermediate-band 
filters from 427~nm to 827~nm, and the CFHT $K_{\rm S}$-band image 
\citep{ilbert09}. In both the COSMOS and AEGIS fields, GALEX photometry in 
the FUV (150~nm) and NUV (225~nm) passbands were added. The NMBS adds six 
filters: $J_{1}$, $J_{2}$, $J_{3}$, $H_{1}$, $H_{2}$, and $K$. Filters 
characteristics of the five medium band filters are given in 
\citet{vandokkum09b}.

The NMBS is an NOAO Survey Program, with 45 nights allocated over three 
semesters (2008A, 2008B, 2009A). An additional 30 nights were allocated 
through a Yale-NOAO time trade. The full details of the reduction, source 
detection, and generation of the photometric catalogs will be described in 
K.~E.~Whitaker et al. (in prep.). In the present analysis, we use a 
$K$-selected catalog based on all the data obtained over the three semesters. 
The AEGIS catalog contains 17 filters and the COSMOS catalog contains 
35 filters ($FUV-8~\mu$m). The images were convolved to the same point-spread 
function (PSF) before performing aperture photometry, so as to limit any 
bandpass-dependent effects. Following previous studies (\citealt{labbe03}; 
\citealt{quadri07}), the photometry was performed with SExtractor in 
relatively small ``color'' apertures which optimize the S/N ratio. Total 
magnitudes in each band were determined from an aperture correction computed 
from the $K$-band. The aperture correction is a combination of the ratio of 
the flux in SExtractor's AUTO aperture to the flux in the color aperture and 
a point-source-based correction for flux outside of the AUTO aperture, 
thereby enabling us to calculate total magnitudes (see, e.g., 
\citealt{labbe03}). The $K$-band completeness limit of the NMBS catalog 
adopted in this work is $K=23.15$~mag. Stars are flagged based on their 
observed $U-J1$ and $J1-K$ colors, where the stellar sequence cleanly 
separates from the bulk of galaxies in color space (see K.~E.~Whitaker et 
al. 2010, in prep., for more details). The total number of objects in the 
$K$-selected sample is 52259, 27520 of which are in the COSMOS field.


\subsection{Photometric Redshifts}\label{subsec-zphot}

Photometric redshifts were determined with the EAZY code \citep{brammer08}, 
using the full $FUV-8~\mu m$ spectral energy distribution (SEDs) ($FUV-K$ 
for objects in the $\sim$50\% of our AEGIS field that does not have Spitzer 
coverage) and $z_{\rm max}=6.0$ (the maximum allowed redshift within EAZY). 
For this study we have used the photometric redshift $z_{\rm peak}$ output 
by EAZY.\footnote{The default template set used in this work consists of 
seven templates: the six templates taken from the optimized template set 
of EAZY, but augmented with emission lines, and a template of a 12.5~Gyr 
old single stellar population. In \S~\ref{sec-dustold} we consider the 
case of an additional template, consisting of a dust-obscured 
($A_{\rm V}=3$~mag), old (1~Gyr) population.}
Publicly available redshifts in the COSMOS and AEGIS fields indicate that 
the redshift errors are very small at $\sigma_{\rm z}/(1+z)<0.02$ at 
$z_{\rm spec}<1.5$. Specifically, the photometric redshifts in COSMOS are in 
excellent agreement with the spectroscopic redshifts made publicly available 
by the zCOSMOS survey \citep{lilly07}, with $\sigma_{\rm z}/(1+z)=0.008$ 
for 1444 objects at $z_{\rm spec}<1.5$. We also find excellent agreement between 
the photometric and spectroscopic redshifts for a larger sample of 2313 
objects at $z_{\rm spec}<1.5$ in AEGIS from the DEEP2 survey \citep{davis03}, 
with $\sigma_{\rm z}/(1+z)=0.017$. Both fields have very few catastrophic 
failures, with only 3\% $>5$~$\sigma$ outliers. Although there are very few 
spectroscopic redshifts of optically-faint $K$-selected galaxies in these 
fields, we note that we found a similarly small scatter 
($\sigma_{\rm z}/(1+z)<0.02$) in a pilot program targeting galaxies from 
the \citet{kriek08} near-IR spectroscopic sample (see 
\citealt{vandokkum09b}). Spectroscopic redshifts also exist for 125 LBGs 
at $z\sim3$ within the AEGIS field from \citet{steidel03}, for which we find 
$\sigma_{\rm z}/(1+z)=0.045$, with 10\% $>5$~$\sigma$ outliers. From the 
formal EAZY errors listed in Table~\ref{tab-1}, we find typical 
$\sigma_{\rm z}/(1+z)=0.04$, perfectly consistent with the scatter 
$\sigma_{\rm z}/(1+z)$ found for LBGs. We conclude that, in the regime of 
interest in this paper, the errors of the photometric redshifts are 
larger than at $z_{\rm spec}<1.5$, as they are dominated by random errors in 
the photometry.

The observed spectral energy distributions (SEDs) with best-fit EAZY 
templates over-plotted are shown in Figure~\ref{fig-eazy1}, together with 
the EAZY redshift probability distributions. As shown in 
Figure~\ref{fig-eazy1}, the medium-band filters $H_{1}$ and $H_{2}$ allow us 
to identify the redshifted Balmer/4000\AA~ breaks within the $H$ band, 
improving the accuracy of the photometric redshift estimates with respect 
to previous analysis with only broad-band photometry.

\begin{figure*}
\epsscale{1}
\plotone{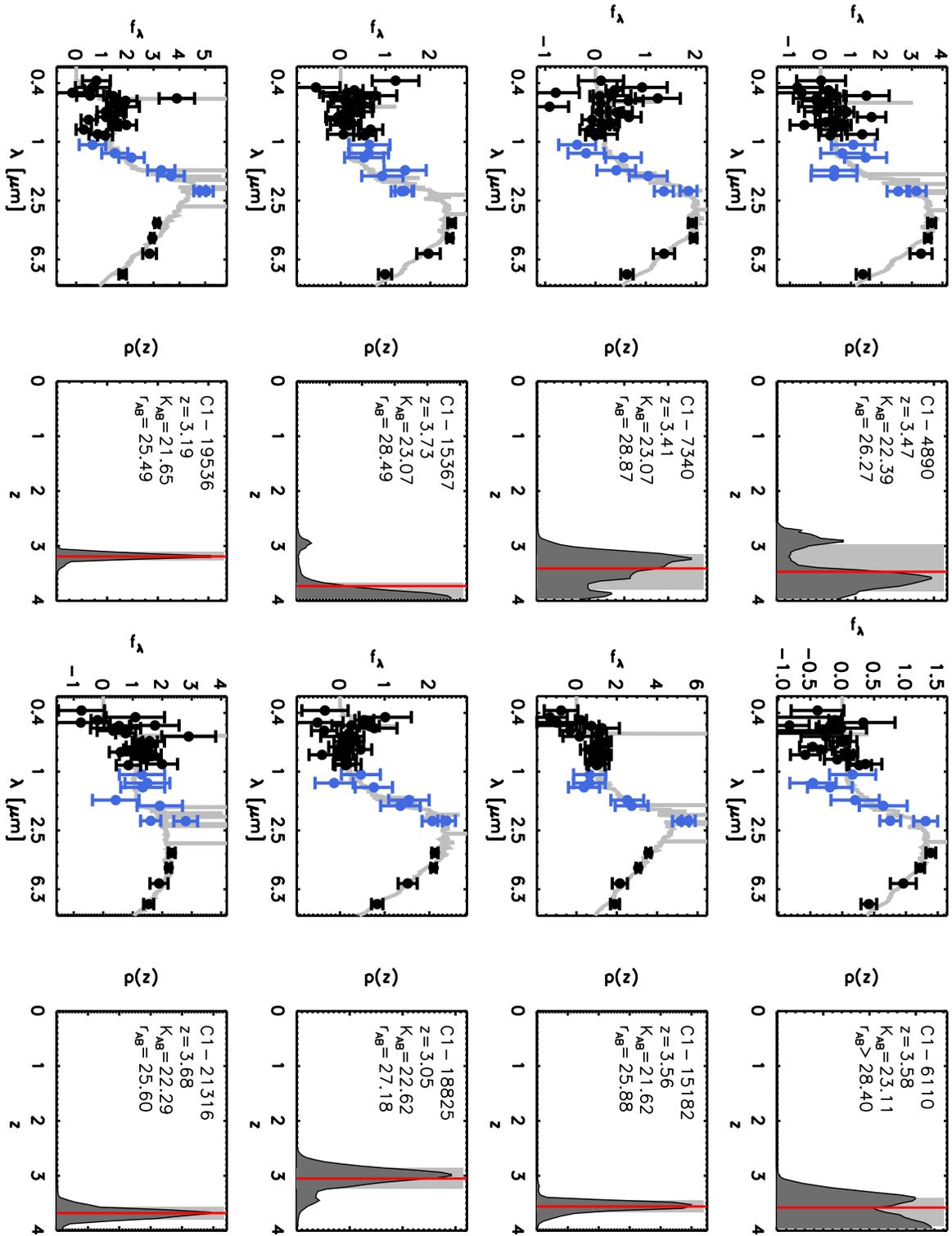}
\caption{Observed SEDs of the mass-selected sample at 
$3.0 \leq z < 4.0$. Filled circles are the observed fluxes, 
in units of 10$^{-19}$~erg~cm$^{-2}$~s$^{-1}$~\AA$^{-1}$, with 
corresponding 1~$\sigma$ errors. The blue symbols are the 
photometric points from NMBS. The solid gray curves represent 
the best-fit EAZY templates. The dark gray filled regions 
represent the EAZY redshift probability functions. The vertical 
red line is the adopted redshift from EAZY ($z_{\rm peak}$, as 
specified in \S~\ref{subsec-zphot}), while the shaded gray 
regions are the 1~$\sigma$ allowed values for the photometric 
redshifts. Also listed are the NMBS identification number, the 
adopted photometric redshift from EAZY, and the observed total 
$K$- and $r$-band magnitudes. \label{fig-eazy1}}
\end{figure*}

\begin{figure*}
\epsscale{0.75}
\plotone{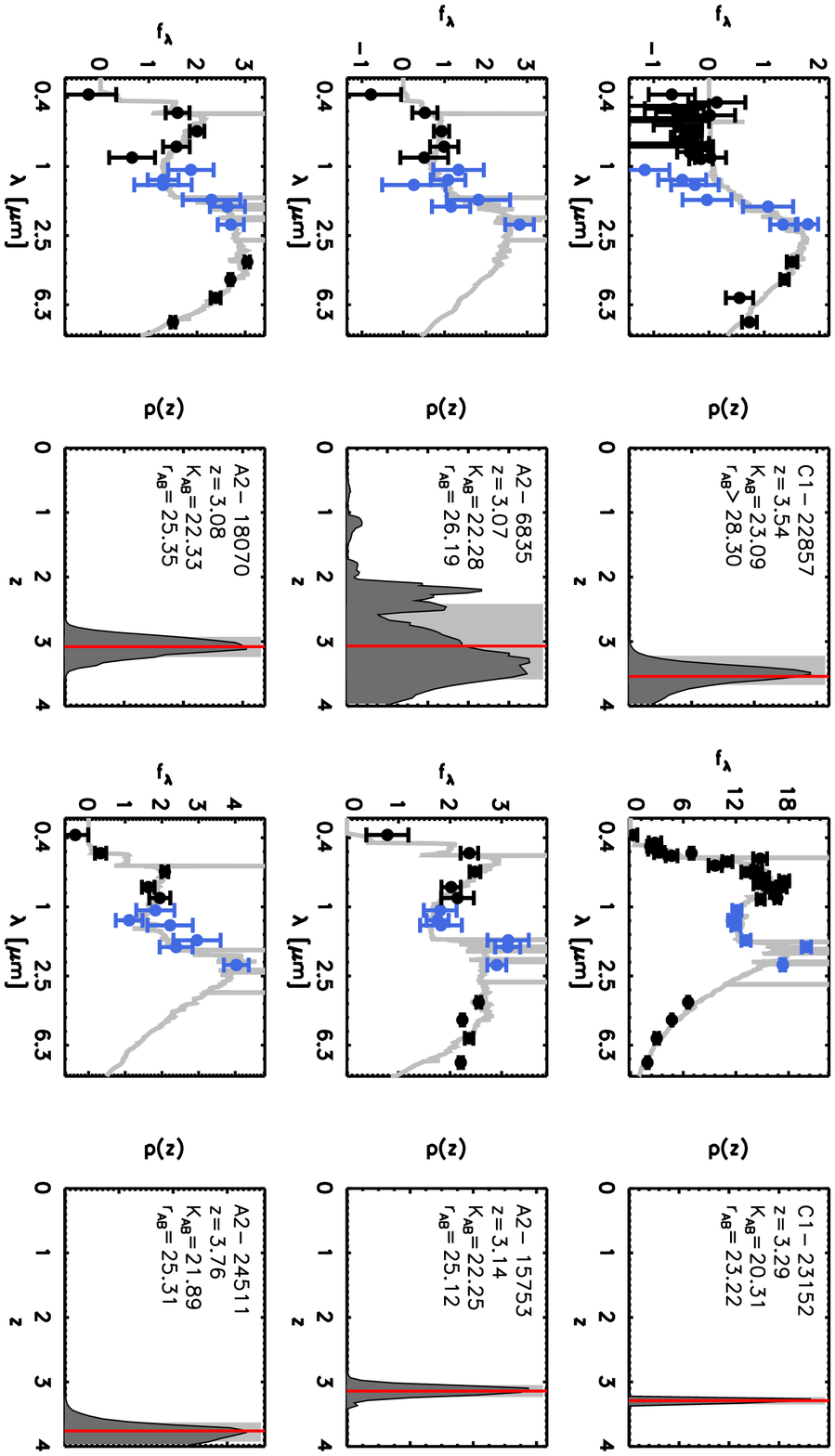}
\caption{Continued from Fig.~\ref{fig-eazy1}}
\end{figure*}

Rest-frame colors were measured using the best-fit EAZY templates, as 
described in \citet{brammer09} and, in particular, in \citet{whitaker10}. 
Briefly, from the best-fit EAZY template, we computed the rest-frame 
$U-V$ colors following the method used by \citet{wolf03} in the COMBO-17 
survey. We used the \citet{maizapellaniz06} filter definitions and used 
the direct template fluxes to determine $U-V$. When using closely-spaced 
medium-band observed filters, the template fluxes are found to be more 
robust than interpolating between observed filters (see Brammer et al. 
2010, in prep.).


\subsection{SED Modeling}\label{subsec-sed}

\begin{figure}
\epsscale{1.1}
\plotone{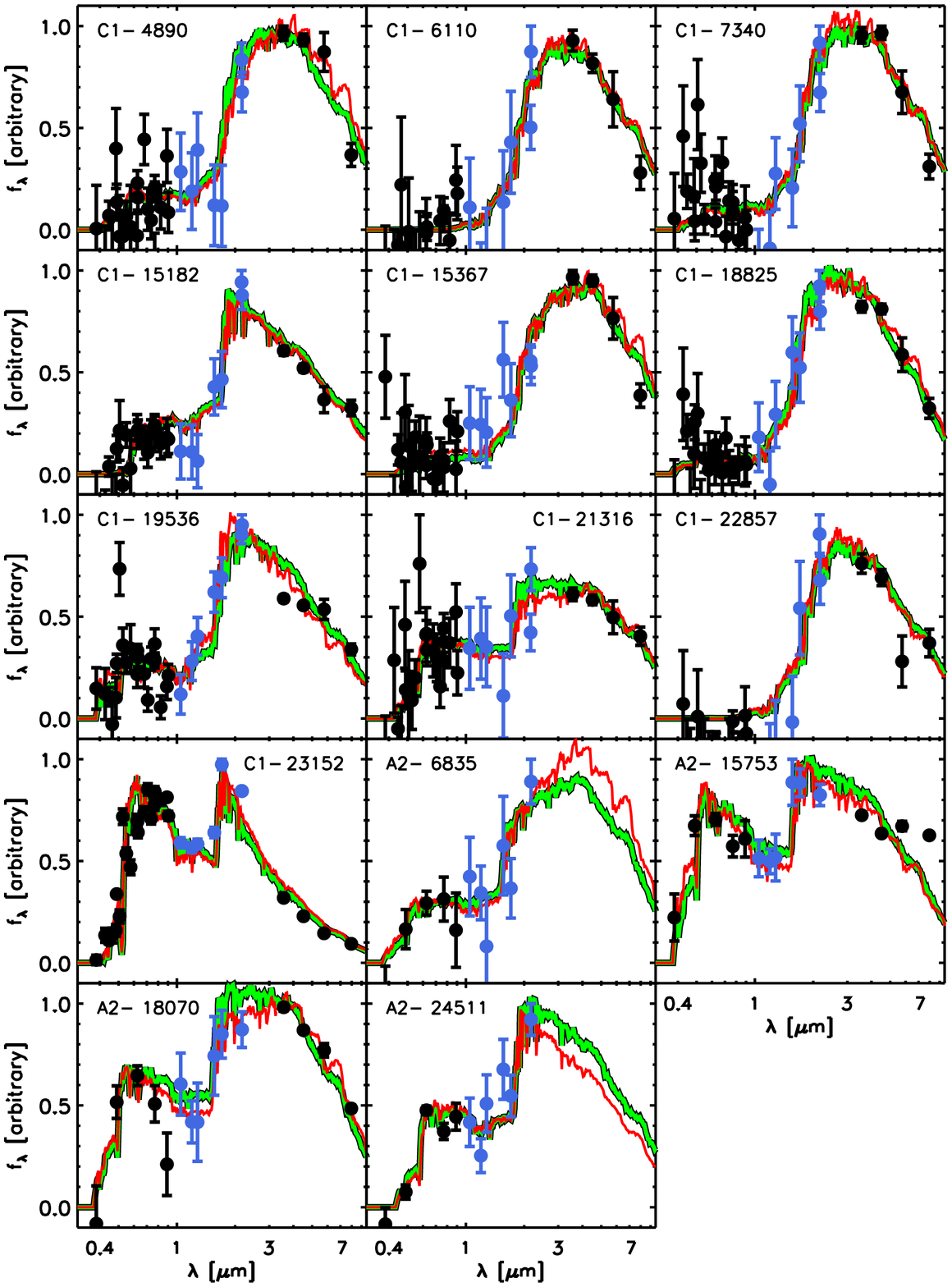}
\caption{Observed SEDs of the mass-selected sample of galaxies at 
$3.0 \leq z < 4.0$. Filled circles are the observed fluxes in 
arbitrary units, with corresponding 1~$\sigma$ errors. The blue 
symbols are the photometric points from NMBS. The solid green curves 
represent the best-fit FAST templates using the stellar population 
synthesis models of \citet{bruzual03}. The solid red curves represent 
the best-fit FAST templates using the stellar population synthesis 
models of \citet{maraston05}. Both stellar population synthesis 
models provide generally good fits to the observed SEDs. 
\label{fig-fast1}}
\end{figure}

Stellar masses and other stellar population parameters were determined 
with FAST \citep{kriek09a}, fixing the redshift to the EAZY output (or the 
spectroscopic redshift when available). For consistency with previously 
published SMF measurements and straightforward comparisons, we assumed the 
default SED-modeling assumptions of \citet{marchesini09}, i.e., stellar 
population synthesis models of \citet{bruzual03}, the \citet{calzetti00} 
reddening law with $A_{\rm V}$ values ranging from 0 to 4 in step of 0.2~mag, 
solar metallicity, pseudo-\citet{kroupa01}\footnote{SED modeling was 
performed using a \citet{salpeter55} IMF with lower and upper mass cutoffs of 
0.1~M$_{\sun}$ and 100~M$_{\sun}$, and the derived stellar masses and star 
formation rates were scaled to a pseudo-\citet{kroupa01} IMF by dividing 
by a factor of 1.6.} initial mass function (IMF), and 
three star formation histories (SFHs): a single stellar population (SSP), 
a constant star formation history (CSF), and an exponentially declining 
SFH with an e-folding timescale of 300~Myr ($\tau_{300}$). In order to 
quantify the systematic uncertainties due to different SED-modeling 
assumptions on the derived stellar population properties (i.e., $M_{\rm star}$, 
age, star formation rate, and $A_{\rm V}$) of the $3.0\leq z < 4.0$ sample, 
we have also assumed the stellar population synthesis models of 
\citet{maraston05} with a \citet{kroupa01} IMF and exponentially declining 
star formation histories with values of the e-folding timescale ranging from 
100~Myr to 10~Gyr in step of 0.2~dex. We refer to \citet{marchesini09} for 
a detailed analysis of the systematic uncertainties on the SMF measurements 
due to the different SED-modeling assumptions. Figure~\ref{fig-fast1} shows 
the observed SEDs of the mass-selected sample at $3.0 \leq z < 4.0$ together 
with the best-fit stellar population models from FAST for our two sets of 
SED-modeling assumptions.

\subsection{The $3 .0 \leq z < 4.0$ Mass-selected Sample}\label{subsec-sample}

We constructed a mass-selected sample of galaxies at $3 .0 \leq z < 4.0$ 
to study their observed and rest-frame properties, as well as to derive 
more accurate measurements of the high-mass end of the SMF of galaxies at 
$3 .0 \leq z < 4.0$. 

\begin{deluxetable*}{lcccccccc}
\centering
\tablecaption{Mass-selected sample of $3.0 \leq z < 4.0$ galaxies\label{tab-1}}
\tablehead{\colhead{ID} & \colhead{$r$} & 
           \colhead{$H$} &  \colhead{$K$} & 
           \colhead{$z$} & \colhead{$\log{M_{\rm star}}$} & 
           \colhead{$SFR$} &  \colhead{$A_{\rm V}$} &  \colhead{$\log{Age}$} \\
                     &  (mag) &  (mag) &  (mag) &  & 
                  (M$_{\sun}$) & (M$_{\sun}$~yr$^{-1}$) & (mag) & (yr) }
\startdata
C1-4890  & 26.27   & 24.90 & 22.39$\pm$0.16 & 3.47$^{+0.34}_{-0.50}$ & 11.70$^{+0.17}_{-0.04}$~(11.67$^{+0.07}_{-0.24}$) & 40.7$^{+188}_{-25.3}$~(33.9$^{+313}_{-20.1}$)   & 1.4$^{+0.6}_{-0.4}$~(1.2$^{+1.2}_{-0.4}$) & 9.1$^{+0.1}_{-0.2}$~(9.2$^{+0.1}_{-1.0}$) \\
C1-6110  & $>28.4$ & 24.91 & 23.11$\pm$0.16 & 3.58$^{+0.32}_{-0.25}$ & 11.41$^{+0.07}_{-0.13}$~(11.30$^{+0.08}_{-0.19}$) & 0.0$^{+0.1}_{-0.0}$~(0.0$^{+0.1}_{-0.0}$)  & 1.0$^{+0.6}_{-0.2}$~(0.8$^{+0.4}_{-0.5}$) & 9.2$^{+0.0}_{-0.3}$~(9.2$^{+0.0}_{-0.1}$) \\
C1-7340  & 28.87   & 24.32 & 23.07$\pm$0.15 & 3.41$^{+0.36}_{-0.27}$ & 11.46$^{+0.07}_{-0.11}$~(11.31$^{+0.07}_{-0.11}$) & 8.3$^{+11.6}_{-6.5}$~(2.3$^{+4.1}_{-2.3}$)   & 1.2$^{+0.4}_{-0.4}$~(0.7$^{+0.5}_{-0.2}$) & 9.2$^{+0.1}_{-0.1}$~(9.2$^{+0.1}_{-0.1}$) \\
C1-15182 & 25.88   & 22.97 & 21.62$\pm$0.09 & 3.56$^{+0.11}_{-0.11}$ & 11.54$^{+0.04}_{-0.05}$~(11.45$^{+0.05}_{-0.09}$) & 0.0$^{+0.1}_{-0.0}$~(0.0$^{+18.2}_{-0.0}$) & 1.6$^{+0.2}_{-0.4}$~(1.3$^{+0.4}_{-0.3}$) & 8.3$^{+0.2}_{-0.1}$~(8.3$^{+0.2}_{-0.2}$) \\
C1-15367 & 28.49   & 23.86 & 23.07$\pm$0.19 & 3.73$^{+0.22}_{-0.06}$ & 11.71$^{+0.05}_{-0.30}$~(11.51$^{+0.12}_{-0.02}$) & 14.8$^{+74.3}_{-13.3}$~(13.2$^{+18.4}_{-9.2}$)  & 1.4$^{+1.2}_{-0.2}$~(1.1$^{+0.4}_{-0.3}$) & 9.2$^{+0.0}_{-1.2}$~(9.1$^{+0.1}_{-0.1}$) \\
C1-18825 & 27.18   & 23.63 & 22.62$\pm$0.12 & 3.05$^{+0.19}_{-0.19}$ & 11.40$^{+0.04}_{-0.01}$~(11.26$^{+0.10}_{-0.22}$) & 1.9$^{+0.2}_{-0.1}$~(2.0$^{+4.0}_{-0.5}$) & 0.8$^{+0.2}_{-0.2}$~(0.8$^{+0.5}_{-0.2}$) & 9.3$^{+0.0}_{-0.1}$~(9.2$^{+0.1}_{-0.2}$) \\
C1-19536 & 25.49   & 22.66 & 21.65$\pm$0.06 & 3.19$^{+0.07}_{-0.08}$ & 11.55$^{+0.03}_{-0.03}$~(11.22$^{+0.09}_{-0.03}$) & 28.2$^{+2.9}_{-1.9}$~(2.9$^{+7.8}_{-0.2}$)   & 1.0$^{+0.2}_{-0.2}$~(0.1$^{+0.5}_{-0.1}$) & 9.1$^{+0.1}_{-0.1}$~(9.0$^{+0.1}_{-0.1}$) \\
C1-21316 & 25.60   & 23.79 & 22.29$\pm$0.16 & 3.68$^{+0.12}_{-0.11}$ & 11.52$^{+0.01}_{-0.61}$~(11.39$^{+0.11}_{-0.58}$) & 316$^{+1233}_{-240}$~(251$^{+1527}_{-179}$)   & 1.8$^{+0.6}_{-0.2}$~(1.6$^{+0.8}_{-0.5}$) & 9.1$^{+0.1}_{-1.4}$~(9.1$^{+0.1}_{-1.5}$) \\
C1-22857 & $>28.3$ & 24.64 & 23.09$\pm$0.19 & 3.54$^{+0.20}_{-0.17}$ & 11.42$^{+0.02}_{-0.08}$~(11.25$^{+0.05}_{-0.04}$) & 0.0$^{+0.1}_{-0.0}$~(0.0$^{+0.1}_{-0.0}$)   & 0.8$^{+0.2}_{-0.2}$~(0.4$^{+0.1}_{-0.3}$) & 9.2$^{+0.1}_{-0.2}$~(9.2$^{+0.1}_{-0.1}$) \\
C1-23152 & 23.22   & 20.96 & 20.31$\pm$0.02 & 3.29$^{+0.06}_{-0.06}$ & 11.42$^{+0.01}_{-0.01}$~(11.37$^{+0.02}_{-0.01}$) & 0.1$^{+1.4}_{-0.1}$~(0.9$^{+1.3}_{-0.9}$)   & 0.8$^{+0.2}_{-0.2}$~(0.7$^{+0.1}_{-0.1}$) & 8.1$^{+0.1}_{-0.1}$~(8.0$^{+0.1}_{-0.1}$) \\
 & & & & & & & & \\
A2-6835  & 26.19   & 23.61 & 22.28$\pm$0.13 & 3.07$^{+0.54}_{-0.70}$ & 11.48$^{+0.24}_{-0.19}$~(11.48$^{+0.38}_{-0.66}$) & 178$^{+385}_{-172}$~(112$^{+935}_{-112}$)   & 2.0$^{+0.2}_{-0.2}$~(1.7$^{+0.2}_{-1.6}$) & 9.3$^{+0.0}_{-0.7}$~(9.3$^{+0.0}_{-1.7}$) \\
A2-15753 & 25.12   & 22.79 & 22.25$\pm$0.06 & 3.14$^{+0.10}_{-0.09}$ & 11.40$^{+0.02}_{-0.09}$~(11.09$^{+0.16}_{-0.06}$) & 148$^{+56.3}_{-13.0}$~(60.3$^{+180}_{-16.6}$)   & 1.4$^{+0.8}_{-0.2}$~(1.0$^{+0.2}_{-0.2}$) & 9.3$^{+0.0}_{-0.2}$~(8.8$^{+0.5}_{-0.1}$) \\
A2-18070 & 25.35   & 23.04 & 22.33$\pm$0.11 & 3.08$^{+0.16}_{-0.15}$ & 11.44$^{+0.03}_{-0.01}$~(11.30$^{+0.12}_{-0.17}$) & 166$^{+7.8}_{-3.9}$~(151$^{+77.7}_{-107}$)   & 1.6$^{+0.6}_{-0.2}$~(1.5$^{+0.2}_{-0.2}$) & 9.3$^{+0.0}_{-0.1}$~(9.2$^{+0.1}_{-0.4}$) \\
A2-24511 & 25.32   & 22.97 & 21.89$\pm$0.09 & 3.76$^{+0.17}_{-0.12}$ & 11.68$^{+0.11}_{-0.10}$~(11.38$^{+0.28}_{-0.36}$) & 170$^{+298}_{-125}$~(97.7$^{+594}_{-97.7}$) & 1.4$^{+0.2}_{-0.2}$~(1.3$^{+0.4}_{-0.6}$) & 8.9$^{+0.3}_{-0.2}$~(8.4$^{+0.4}_{-0.7}$) \\
\enddata
\tablecomments{``C1'' and ``A2'' refer to the COSMOS and AEGIS fields, 
respectively. 
The listed redshift is the adopted best-fit EAZY redshift $z_{\rm peak}$.
The stellar population parameters were derived using a pseudo-\citet{kroupa01} 
IMF, \citet{bruzual03} stellar population synthesis models, and a 
\citet{calzetti00} extinction law (see \S~\ref{subsec-sed}). Quoted errors 
are the 1~$\sigma$ confidence intervals output by FAST (see \citealt{kriek09a} 
for a detailed description of the adopted method in FAST to estimate 
confidence intervals). The values in parenthesis correspond to the best-fit 
stellar population parameters assuming a \citet{kroupa01} IMF, 
\citet{maraston05} stellar population synthesis models, exponentially 
declining SFHs, and a \citet{calzetti00} extinction law (see 
\S~\ref{subsec-sed}). $H$-band magnitudes are derived by averaging the 
$H1$- and $H2$-band magnitudes.} 
\end{deluxetable*}

The redshift-dependent completeness limit in stellar mass has been 
estimated following the approach described in detail in \citet{marchesini09}, 
which exploits the availability of samples with different depths. The 
completeness of a sample is estimated empirically from the available deeper 
samples, namely, the FIRES (\citealt{labbe03}; \citealt{forster06}) and the 
FIREWORKS \citep{wuyts08} catalogs. Briefly, to estimate the 
redshift-dependent stellar mass completeness limit of the NMBS sample, we 
first selected galaxies belonging to the available deeper samples. Then, we 
scaled their fluxes and stellar masses to match the $K$-band completeness 
limit of the NMBS sample. The upper envelope of points in the 
($M_{\rm star,scaled}-z$) space, encompassing 95\% of the points, represents 
the most massive galaxies at the considered flux limit ($K=23.15$ for the 
NMBS catalog adopted in this work), and so provides a redshift-dependent 
stellar mass completeness limit for the NMBS sample. We refer to 
\citet{marchesini09} for a detailed description of this method. The 
resulting completeness in mass of the NMBS catalog used in this work is 
$M_{\rm star} = 10^{11.40}$~M$_{\sun}$ $\approx 2.5\times10^{11}$~M$_{\sun}$ over 
the targeted redshift range $3 .0 \leq z < 4.0$.

The resulting mass-selected sample of galaxies at $3 .0 \leq z < 4.0$ 
contains 14 sources with $M_{\rm star} \geq 10^{11.40}$~M$_{\sun}$ (10 
from the COSMOS field and 4 from the AEGIS field) over an effective 
area of 0.44~square degrees. The sample is listed in Table~\ref{tab-1}, 
along with the observed $r$-, $H$-, and $K$-band total magnitudes, 
adopted EAZY best-fit redshifts and 1~$\sigma$ errors, and FAST best-fit 
$M_{\rm star}$, SFR, $A_{\rm V}$, and age with corresponding 1~$\sigma$ errors. 

As shown in Table~\ref{tab-1}, the typical random error on the estimated 
stellar masses of the mass-selected sample is $\sim0.1$~dex for the default 
set of SED-modeling assumptions (which uses the \citealt{bruzual03} models), 
and $\sim0.16$~dex for the other set (which adopts the \citealt{maraston05} 
models). These errors are in good agreement with the errors on stellar 
mass due to photometric redshift uncertainties estimated by \citet{taylor09}, 
with a typical error on the stellar mass of $\sim0.1$~dex for photometric 
redshift errors of $\sigma_{\rm z}/(1+z)=0.035$ at $z<1.5$. As shown by 
\citet{taylor09}, in a photometric redshift survey, the stellar mass 
estimates are relatively robust to random photometric redshift errors, due 
to the similar (but opposite) systematic effects on luminosities and stellar 
mass-to-light ratios caused by random photometric redshift errors.

\begin{figure*}
\epsscale{1.1}
\plotone{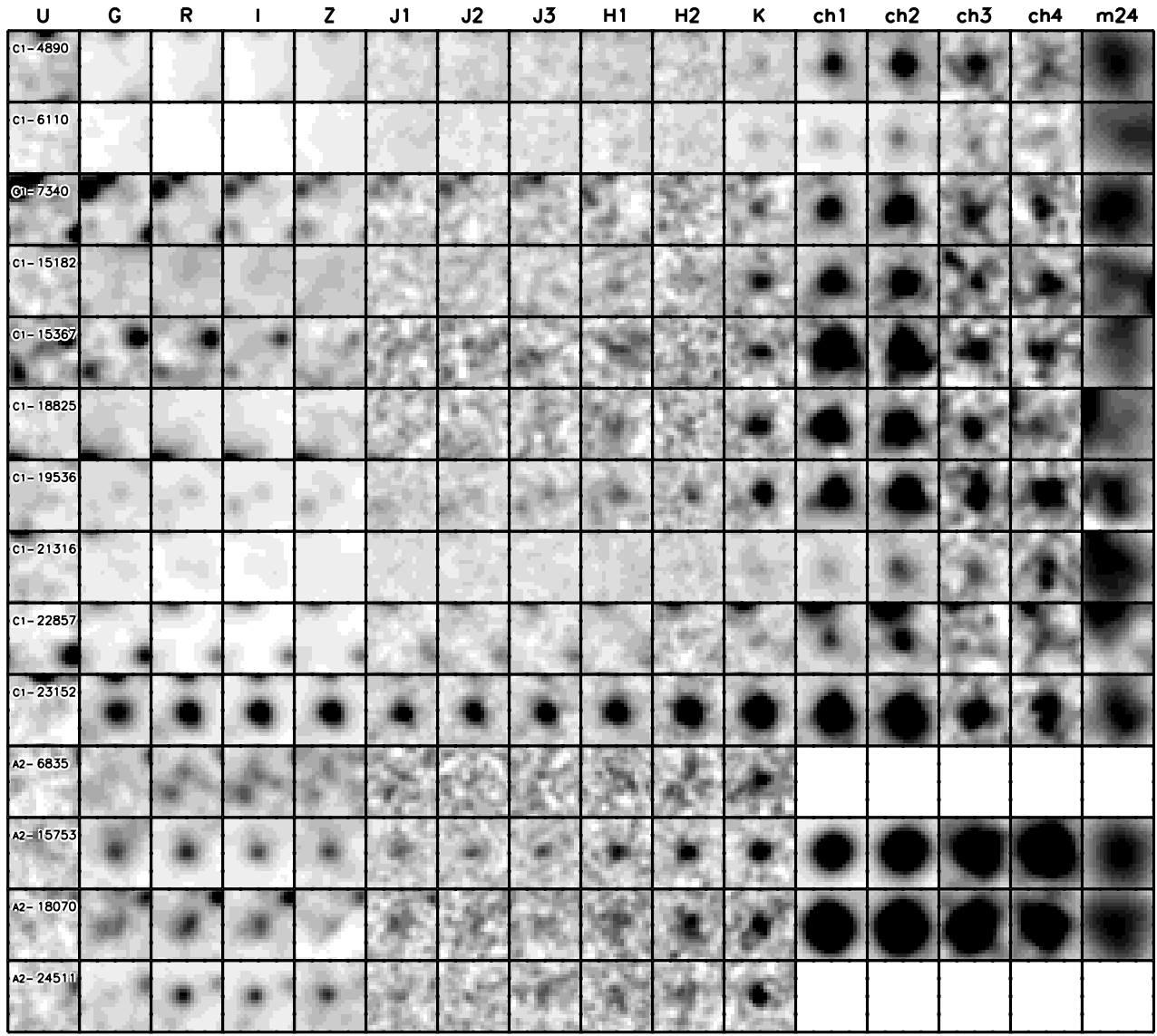}
\caption{Images of the mass-selected sample of galaxies at 
$3.0 \leq z < 4.0$. From left to right, the columns show the 
CFHTLS $u$, $g$, $r$, $i$, and $z$ images, the NMBS $J1$, $J2$, 
$J3$, $H1$, $H2$, and $K$ images, the {\it Spitzer}-IRAC 
3.6~$\mu$m, 4.5~$\mu$m, 5.8~$\mu$m, and 8.0~$\mu$m images, and 
the {\it Spitzer}-MIPS 24~$\mu$m image. Each cutout is 
6$^{\prime \prime} \times$6$^{\prime \prime}$ on a side.
\label{fig-stamps}}
\end{figure*}

Figure~\ref{fig-stamps} shows the images of the mass-selected sample 
at $3.0 \leq z < 4.0$ in the different filters, from the $u$-band to the 
$24$~$\mu$m {\it Spitzer}-MIPS channel. 

In order to exclude contamination of the mass-selected sample due to 
blending, we have also inspected the higher-spatial resolution images 
available over COSMOS and AEGIS. For COSMOS, we have used the CFHT-WIRCAM 
$K_{\rm S}$-band image (FWHM$\sim$0.8$^{\prime \prime}$) and the HST-ACS 
$I_{\rm F814W}$-band images.\footnote{Available at 
\url{http://irsa.ipac.caltech.edu/data/COSMOS/datasets.html}.} For AEGIS, 
we have used the HST-ACS $I_{\rm F814W}$-band images.\footnote{Available at 
\url{http://aegis.ucolick.org/acs\_data\_descrip.html}.} In the HST images, 
only C1-19536, C1-23152, and A2-15753 are detected, whereas the other 
sources do not show any obvious detection. All three detected sources 
appear to be resolved in the ACS images, indicating that the $I$-band 
fluxes are not dominated by a point-like component. Specifically, C1-19536 
and A2-15753 are extended and quite elongated. All the sources but C1-23152 
appear very isolated in the ACS images, consistent with the ground-based 
images. The HST-ACS $I_{\rm F814W}$-band image of C1-23152 reveals two fainter 
knots at a distance of $\sim1.1^{\prime \prime}$. The two knots contribute about 
11\% to its total flux in the ACS image. Inspection of the $K_{\rm S}$-band 
image over COSMOS reveals that all selected massive galaxies are single, 
isolated objects, including C1-23152, showing no obvious signature of the 
two knots. If the photometry of C1-23152 is equally affected by the two 
knots at all wavelengths, the shape of its SED would not be affected, and 
the resulting stellar mass would be smaller by $\sim0.05$~dex, not changing 
the results of this paper. On the contrary, if the contribution of the two 
knots changes as a function of wavelength, the observed SED would change 
accordingly, making it harder to predict how the derived stellar mass would 
be affected. A rough estimate of this effect was derived by re-fitting the 
observed SED of C1-23152 after assuming that only the optical fluxes are 
affected by the two knots. The resulting stellar mass is only $\sim0.03$~dex 
smaller than that estimated with the current photometry, implying that 
the derived stellar mass for C1-23152 is not likely to be significantly 
biased by the two knots. We therefore conclude that none of the observed 
objects seem to be affected by blending issues, which might have resulted 
in systematically biased stellar mass estimates. Higher spatial-resolution 
NIR imaging is however required to confirm this.

Finally, we note that no a priori exclusion of active galactic nuclei 
(AGNs) has been performed in our mass-selected sample.


\section{PROPERTIES OF VERY MASSIVE GALAXIES AT $3.0 \leq z < 4.0$} \label{sec-prop}

We use our mass-selected sample of 14 galaxies to determine the median and 
dispersion in observed and rest-frame properties of the most massive galaxies 
($M_{\rm star} \gtrsim 2.5 \times 10^{11}$~M$_{\sun}$) at $3.0 \leq z < 4.0$. 
Table~\ref{tab-2} lists the median and 25th/75th percentiles of the 
distributions of observed $r$-band magnitude and $H-K$ color; the rest-frame 
$V$-band magnitude and $U-V$ color; and rest-frame UV slopes, parametrized 
by $F_{\rm \lambda} \propto \lambda^{\beta}$. Rest-frame UV slopes $\beta$ were 
determined from the best-fitting SEDs, following the \citet{calzetti94} 
method of fitting to the 10 rest-frame UV bins defined by those authors. 

\begin{figure}
\epsscale{1.1}
\plotone{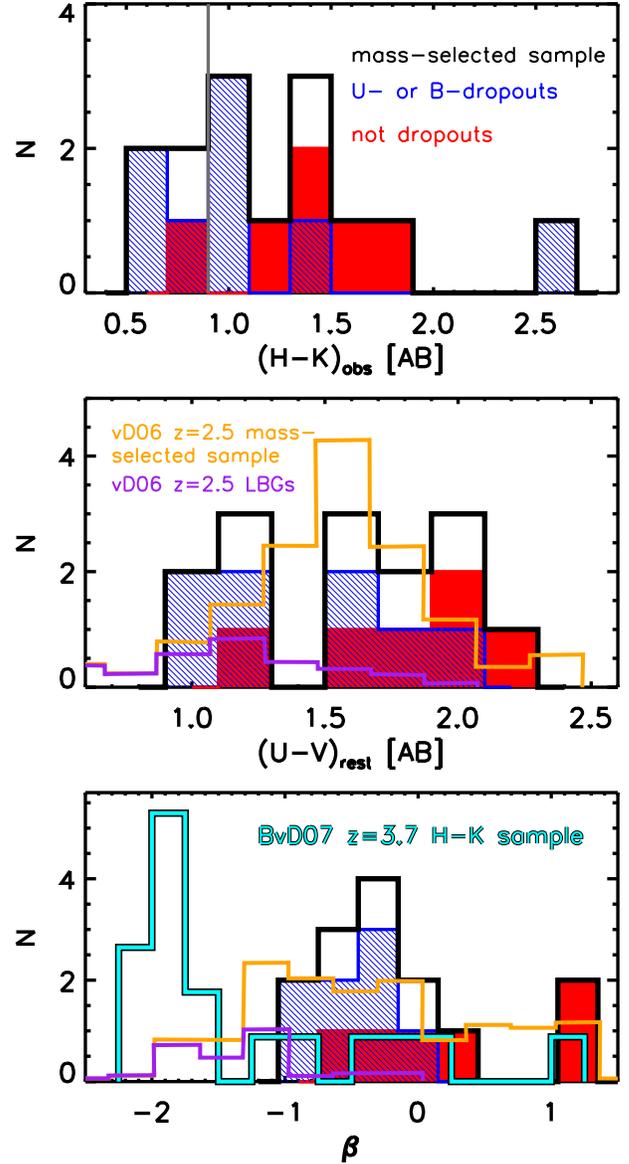}
\caption{{\it Top panel:} Distribution of observed $H-K$ colors of the 
mass-selected sample of galaxies at $3.0 \leq z < 4.0$ (solid black line); 
the blue hatched area represents the distribution for those galaxies 
that would be selected as either $U$- or $B$-dropout galaxies based on 
their observed optical colors; the red filled area represents the 
distribution for those galaxies that would not be selected as either $U$- 
or $B$-dropout galaxies based on their observed optical colors; the 
vertical gray line represents the $H-K$ criterion adopted in 
\citet{brammer07} to select galaxies at $z\sim3.7$. 
{\it Middle panel:} Distribution of rest-frame $U-V$ colors of the 
mass-selected sample; colors as in the top panel; the orange solid 
line represents the distribution of rest-frame $U-V$ colors of the 
mass-selected sample at $2<z<3$ from \citet{vandokkum06}; the purple 
solid line represents the distribution of rest-frame $U-V$ colors of 
the galaxies in the mass-selected sample at $2<z<3$ from \citet{vandokkum06} 
that would be selected as LBGs.
{\it Bottom panel:} Distribution of rest-frame UV slopes of the 
mass-selected sample; colors as in the middle panel; the solid cyan line 
represents the distribution of UV slopes of the $H-K$ selected sample 
at $z\sim3.7$ of \citet{brammer07}.
\label{fig-histos}}
\end{figure}

Figure~\ref{fig-histos} shows the distributions of the observed $H-K$ color 
(top panel), rest-frame $U-V$ color (middle panel), and rest-frame UV 
slopes of the mass-selected sample (bottom panel). For comparison, we have 
also plotted 1) the distribution of rest-frame $U-V$ colors and UV slopes of 
a mass-selected sample of galaxies at $2<z<3$ with 
$M_{\rm star}>6\times10^{10}$~M$_{\sun}$ from \citet{vandokkum06} (orange 
histogram); 2) the distribution of rest-frame $U-V$ colors and UV slopes of 
the galaxies that would be selected as LBGs 
from the sample of \citet{vandokkum06} (purple histogram); and 3) the 
distribution of rest-frame UV slopes of a $z\sim3.7$ sample of galaxies 
from \citet{brammer07} selected with the color criterion $H-K>0.9$ to have 
prominent Balmer/4000~\AA~ breaks between the $H$ and $K$ bands (cyan 
histogram).

\begin{deluxetable}{lccc}
\centering
\tablecaption{Observed and rest-frame properties of the 
$3.0 \leq z < 4.0$ mass-selected sample\label{tab-2}}
\tablehead{\colhead{Quantity} & \colhead{25\%} & 
           \colhead{Median} & 
           \colhead{75\%} }
\startdata
$r_{\rm tot}$~(obs)  &  25.3 &  26.1 & 27.8  \\
$H-K$~(obs)       &  0.75 &  1.16 & 1.42  \\
$V_{\rm tot}$~(rest) & $-$24.2 & $-$23.5 & $-$23.3 \\
$U-V$~(rest)      &  1.26 &  1.64 & 1.90  \\
$\beta$~(rest)    & $-$0.56 & $-$0.36 & 0.07  \\
\enddata
\end{deluxetable}

The typical very massive galaxy at $3.0 \leq z < 4.0$ (median stellar 
mass $\langle M_{\rm star}\rangle \sim 3 \times 10^{11}$~M$_{\sun}$) is red 
and faint in the observer's optical, with $\langle r_{\rm tot}\rangle=26.1$. 
Most galaxies (10 out of 14) would be selected as $H-K$ 
galaxies with $H-K>0.9$ \citep{brammer07}. Of the four galaxies with 
$H-K<0.9$, only two galaxies have $H-K$ color significantly smaller than 
the $H-K$ criterion. This highlights the efficiency of this color technique 
in selecting galaxies at $z>3$ with prominent breaks in the rest-frame 
optical, although the fraction of interlopers selected by this color 
technique remains uncertain.

From Table~\ref{tab-1}, 40\%-60\% of the very massive galaxies are 
characterized by ages consistent with the age of the universe at the 
targeted redshifts ($\sim1.6-2.1\times10^{9}$~yr). About 30\% of the very 
massive galaxies, namely C1-6110, C1-15182, C1-22857, and C1-23152 
have SFR estimates from SED modeling consistent with no star formation 
activity to within 1~$\sigma$, independent of the specific SED-modeling 
assumptions adopted in FAST. Of the remaining galaxies, 4 have very large 
SFRs, of the order of a few hundreds solar masses per year. We stress that 
the estimated ages and star formation rates from SED modeling are quite 
uncertain, even with the high-quality dataset used in this work (e.g., 
\citealt{muzzin09}). 

\subsection{Rest-frame UV}

The rest-frame $U-V$ colors range from $U-V=1.01$, typical of nearby 
irregular galaxies, to $U-V=2.2$, typical of local elliptical galaxies 
(e.g., \citealt{fukugita95}). The median $\langle U-V \rangle=1.64$ is 
similar to local Sb spiral galaxies. As shown in the middle panel of 
Figure~\ref{fig-histos}, the mass-selected sample of 
\citet{vandokkum06} at $z=2.5$, which is complete in stellar mass down 
to $\sim6\times10^{10}$~M$_{\sun}$ (a factor of $\sim5$ less than our sample), 
encompasses the range in $U-V$ colors spanned by our $z=3.5$ sample, with 
a median $U-V$ color bluer by $\sim0.1$~mag with respect to our 
mass-selected sample.

The median UV slope $\beta$ is $\langle \beta \rangle=-0.36$, indicating 
a relatively flat spectrum in $F_{\lambda}$. The distribution of $\beta$, 
ranging from $\beta=-0.95$ to $\beta=1.10$, is broadly consistent with the 
distribution of massive galaxies at $2<z<3$ from \citet{vandokkum06}.
As shown in the bottom panel of Figure~\ref{fig-histos}, the distribution 
of $\beta$ is instead very different from the distribution seen for 
$H-K>0.9$ galaxies at $z\sim3.7$, which shows a peak at 
$\beta \sim -2$ \citep{brammer07}. The observed distribution of $\beta$ 
for the mass-selected sample at $3.0 \leq z < 4.0$ is also very different 
from the distributions seen for UV-selected galaxies at $z\sim2.5$ and 
$z\sim4$, which peak at $\beta \sim -1.6$ and $\beta \sim -1.8$, 
respectively (e.g., \citealt{adelberger00}; \citealt{ouchi04}; 
\citealt{hathi08}; \citealt{bouwens09}).

\begin{figure}
\epsscale{1.1}
\plotone{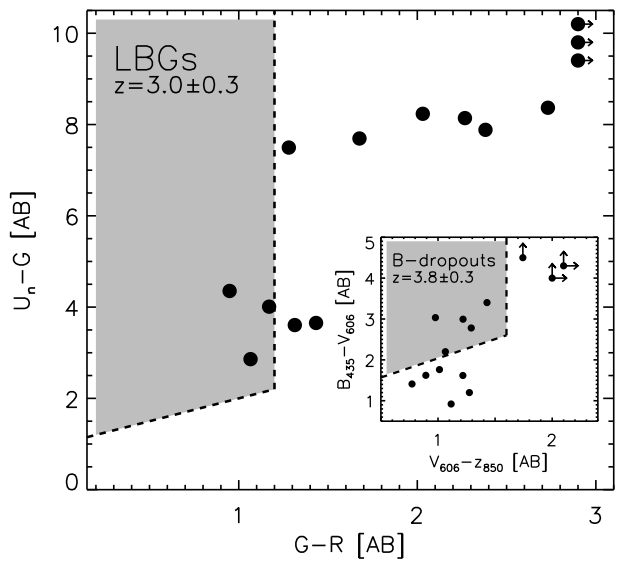}
\caption{Color selection of LBGs at $z\sim3.0\pm0.3$ \citep{steidel03} 
and, in the insert, of $B$-dropout galaxies at $z\sim3.8\pm0.3$ 
\citep{bouwens09}. Objects falling in the gray shaded regions would be 
selected as LBGs or $B$-dropout galaxies. Out of the 14 very massive 
galaxies at $3.0 \leq z < 4.0$, 3 would be selected as LBGs and 5 as 
B-dropout galaxies based on their colors. However, only 4 out of 8 
of the dropout galaxies have $r \leq 25.5$. 
\label{fig-dropouts}}
\end{figure}

The intrinsically different rest-frame UV properties of the mass-selected 
sample at $3.0 \leq z < 4.0$ studied in this work and the typical 
UV-selected galaxies at these redshifts (i.e., $U$- and $B$-dropout 
galaxies) is also clear from Figure~\ref{fig-dropouts}, which shows the 
location of the massive galaxies at $3.0 \leq z < 4.0$ in the $U_{\rm n}GR$ 
and $B_{\rm 435}V_{\rm 606}z_{\rm 850}$ diagrams commonly used to select 
$U$-dropout galaxies (i.e., LBGs; \citealt{steidel03}) and $B$-dropout 
galaxies (\citealt{giavalisco04}; \citealt{bouwens09}), respectively. 
The colors plotted in Figure~\ref{fig-dropouts} are synthetic colors 
integrated from the best-fit FAST templates. About 57\% of the galaxies 
in the $3.0 \leq z < 4.0$ mass-selected sample have colors that satisfy 
either the $U$- or the $B$-dropout color criteria (gray shaded area in 
Figure~\ref{fig-dropouts}). Of these, three would be selected as 
$U$-dropouts, and five as $B$-dropouts, based on their observed optical 
colors. However, $\sim$50\% of these UV-selected galaxies are fainter 
than $r_{\rm tot}=25.5$, which is the observed optical limit of typical 
spectroscopic samples of LBGs. While the $r_{\rm tot}=25.5$ cut is not 
relevant to the inclusion of our galaxies in the photometric window, it 
is relevant when considering our objects in the context of pre-existing 
LBG surveys.

\begin{figure}
\epsscale{1.1}
\plotone{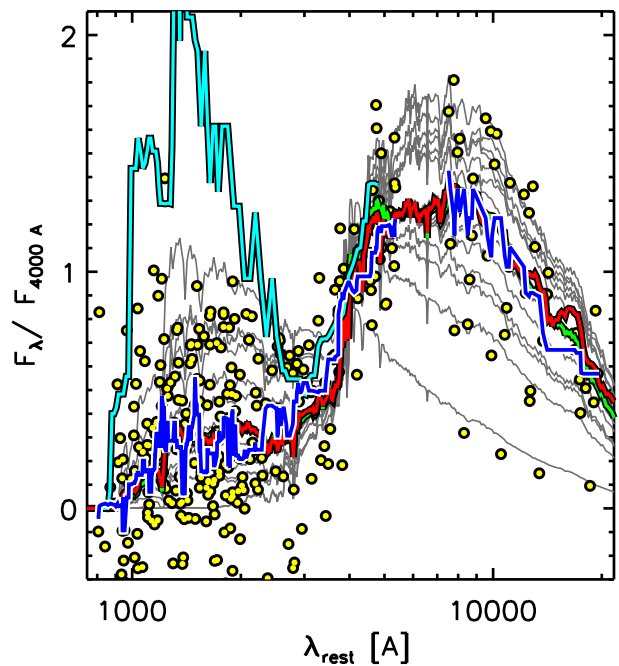}
\caption{Rest-frame SEDs of the mass-selected sample at 
$3.0 \leq z < 4.0$ (yellow filled circles). The SEDs are 
normalized to the flux at $\lambda_{\rm rest}=4000$~\AA. 
The solid blue curve represents the running median of the 
15 neighboring points. The best-fit FAST templates 
adopting the \citet{bruzual03} models are shown as dark 
gray solid curves. The green and red solid curves represent 
the median best-fit FAST templates adopting the 
\citet{bruzual03} and the \citet{maraston05} models, 
respectively. The cyan solid curve represents the median 
SED of the $H-K$ galaxies at $z\sim3.7$ from \citet{brammer07}. 
\label{fig-restframe}}
\end{figure}

The rest-frame SEDs of the mass-selected sample at $3.0 \leq z < 4.0$ are 
shown in Figure~\ref{fig-restframe}, together with the median rest-frame 
SED from the data (solid blue curve) and the median best-fit templates 
from FAST (green and red solid curves). Figure~\ref{fig-restframe} clearly 
shows the strongly-suppressed emission and the flatness of the spectrum in 
$F_{\lambda}$ in the rest-frame UV, as well as the strong Balmer/4000~\AA~breaks 
in the rest-frame optical for the typical very massive galaxy at $z \sim 3.5$. 
Also plotted is the median rest-frame SED of the $H-K$-selected sample at 
$z\sim3.7$ from \citet{brammer07}. Figure~\ref{fig-restframe} clearly shows 
that the rest-frame optical SED of the $H-K$-selected sample and our 
mass-selected sample are very similar, characterized by strong rest-frame 
optical breaks. In contrast, their rest-frame UV SEDs are very different. The $H-K$-selected galaxies are characterized by very 
blue rest-frame UV-optical colors. On the contrary, our mass-selected 
galaxies are generally red also in the rest-frame UV.

The significant differences in observed and rest-frame properties 
between the $H-K$-selected galaxies at $z\sim3.7$ from \citep{brammer07} 
and the very massive galaxies at $3.0 \leq z < 4.0$ selected in our 
study are very interesting, as most of our galaxies would also be selected 
as $H-K$ galaxies. The simplest explanation for the observed differences 
is the very different regime in stellar masses probed by the two samples. 
The $H-K$ galaxies from \citet{brammer07} were selected from the FIRES 
survey over an area of $\sim31$~arcmin$^{2}$, with a median stellar mass a 
factor of $\sim15$ smaller than the median stellar mass of our mass-selected 
sample. The lack of very massive galaxies in FIRES is simply caused by its 
very small surveyed area, as an area of $>100$~arcmin$^{2}$ is required to 
find one single object as massive as our mass-selected galaxies. The 
different stellar mass regime probed by the two samples suggests that 
lower-mass galaxies are characterized by much bluer rest-frame UV-optical 
colors than the most massive galaxies at these redshifts. This is also 
supported by the different rest-frame UV properties between our 
mass-selected sample and typical UV-selected $B$- and $V$-dropout 
samples, which have stellar masses in the range $10^{9}-10^{11}$~M$_{\sun}$ 
(e.g, \citealt{shapley01}; \citealt{magdis10}).
From the SED modeling analysis, it appears that the 
differences in the rest-frame UV properties between the $H-K$-selected 
galaxies and our mass-selected sample could be primarily due to the different 
amount of dust in the two samples, as their typical stellar ages 
($\sim1-1.5$~Gyr) are broadly consistent within the uncertainties. More 
specifically, the $H-K$-selected galaxies from \citet{brammer07} are 
characterized by a median dust extinction of 
$\langle A_{\rm V}\rangle \sim 0.2$, a factor of 
about $\sim7$ smaller than the median extinction of our mass-selected sample 
($\langle A_{\rm V} \rangle=1.4$~mag). Indeed, significant amount of dust seems 
to be quite ubiquitous in massive galaxies at $3.0 \leq z < 4.0$. The 
amount of dust in our mass-selected sample is significantly higher than the 
amount of dust in local massive galaxies and in massive galaxies at 
$z\sim1-2$ ($A_{\rm V}\approx0.2-0.3$~mag, e.g., \citealt{whitaker10}), 
further contributing to suppressing the rest-frame UV light in these 
massive galaxies.\footnote{We note that performing the SED-modeling forcing 
$A_{\rm V}=0$ results in stellar masses smaller by $0-0.2$~dex. However, the 
resulting $\chi^{2}$ of the best-fit models are significantly worse than the 
$\chi^{2}$ values corresponding to the modeling allowing for dust. Moreover, the 
MIPS detections strongly suggest the presence of significant amount of dust. 
Therefore, we conclude that the dust-free assumption is not a realistic 
assumption.}

\subsection{{\it Spitzer}-MIPS 24~$\mu$m data}

We have used the publicly available observations at 24~$\mu$m from 
{\it Spitzer}-MIPS provided by S-COSMOS and 
FIDEL\footnote{\url{http://irsa.ipac.caltech.edu/data/SPITZER/FIDEL/}} to 
further constrain the activity in the most massive galaxies at 
$3.0 \leq z < 4.0$. The measured 24~$\mu$m fluxes with the corresponding 
1~$\sigma$ errors are listed in Table~\ref{tab-3}. For two galaxies over 
AEGIS, namely A2-6835 and A2-24511, no 
MIPS-24~$\mu$m data are available. The 24~$\mu$m flux of C1-18825 is not 
reliable due to blending issues from a nearby very bright 24~$\mu$m source. 
The MIPS cutouts are shown in Figures~\ref{fig-stamps}.

In the sample of 11 massive galaxies at $3.0 \leq z < 4.0$ with MIPS 
coverage and no blending issues, $\sim$80\% have a MIPS 24~$\mu$m fluxes 
significant at $>3$~$\sigma$. This is broadly consistent with the high 
fraction of MIPS-detected sources in the sample of IRAC-selected massive 
($M_{\rm star}\sim10^{10}-10^{11}$~M$_{\sun}$) galaxies at $z>3.5$ over 
GOODS-North \citep{mancini09}. The fraction of MIPS detected massive 
galaxies in our sample increases up to $\sim$90\% for a $>1$~$\sigma$ 
detection. Only C1-22857 is undetected at 24~$\mu$m. 

\begin{deluxetable}{lccc}
\centering
\tablecaption{{\it Spitzer}-MIPS 24~$\mu$m fluxes and 
derived properties of the $3.0 \leq z < 4.0$ mass-selected 
sample\label{tab-3}}
\tablehead{\colhead{ID}        & \colhead{$S_{\rm 24}$} &
           \colhead{$L_{\rm IR}$} & \colhead{SFR}       \\
                  & \colhead{[$\mu$Jy]} &
           \colhead{[10$^{13}$~L$_{\sun}$]}  & \colhead{[M$_{\sun}$~yr$^{-1}$]}}
\startdata
C1-4890  & 206.2$\pm$32.8 & 3.3$\pm$0.5$(^{+2.1}_{-2.3})$ & 3611$\pm$574$(^{+2280}_{-2500})$   \\
C1-6110  & 116.4$\pm$28.0 & 2.3$\pm$0.5$(^{+1.1}_{-0.9})$ & 2458$\pm$591$(^{+1142}_{-1030})$ \\
C1-7340  & 133.0$\pm$28.9 & 1.9$\pm$0.4$(^{+1.5}_{-0.9})$ & 2078$\pm$451$(^{+1682}_{-1025})$  \\
C1-15182 &  78.6$\pm$25.7 & 1.5$\pm$0.5$(^{+0.6}_{-0.6})$ & 1614$\pm$528$(^{+694}_{-663})$  \\
C1-15367 & 167.3$\pm$30.8 & 3.9$\pm$0.7$(^{+1.2}_{-0.8})$ & 4275$\pm$787$(^{+1226}_{-909})$  \\
C1-18825$^{a}$ & blended        & \nodata           & \nodata            \\
C1-19536 &  61.6$\pm$24.6 & 0.5$\pm$0.2$(^{+0.3}_{-0.2})$ &  584$\pm$233$(^{+300}_{-282})$  \\
C1-21316 & 177.1$\pm$31.3 & 4.0$\pm$0.7$(^{+1.0}_{-1.0})$ & 4330$\pm$765$(^{+1064}_{-1095})$  \\
C1-22857 &        $<$19.9 & $<$0.4                      & $<$395  \\
C1-23152 & 110.5$\pm$27.6 & 1.2$\pm$0.3$(^{+0.4}_{-0.4})$ & 1331$\pm$333$(^{+424}_{-408})$  \\
A2-6835$^{b}$ & \nodata       & \nodata           & \nodata              \\
A2-15753 & 165.7$\pm$22.8 & 1.3$\pm$0.3$(^{+0.5}_{-0.5})$ & 1371$\pm$378$(^{+557}_{-476})$  \\
A2-18070 & 127.0$\pm$21.3 & 0.9$\pm$0.3$(^{+0.5}_{-0.4})$ &  918$\pm$308$(^{+582}_{-420})$  \\
A2-24511$^{b}$ & \nodata       & \nodata           & \nodata              \\
\enddata
\tablecomments{$^{a}$no reliable MIPS~24$\mu$m flux could be obtained 
for C1-18825 due to blending issues; $^{b}$no MIPS~24$\mu$m data 
available for A2-6835 and A2-23152. The errors listed for $L_{\rm IR}$ 
and SFR are computed using just the 24~$\mu$m photometric errors 
(values not in parenthesis) and the combination of the 24~$\mu$m 
photometric errors and the photometric redshift errors (values in 
parenthesis).}
\end{deluxetable}

For the redshift range targeted here, the 24~$\mu$m band probes the 
rest-frame wavelengths from $\sim4.8 \mu$m to $\sim7.1 \mu$m, which 
includes the 5.27$\mu$m, 5.7$\mu$m, and 6.2$\mu$m emission features 
from polycyclic aromatic hydrocarbons (PAHs) \citep{draine07}. MIR 
emission at these wavelengths could arise from warm/hot dust and PAH 
molecules, heated by either dust-enshrouded star formation or AGN.
The 24~$\mu$m emission is widely used to estimate the SFRs in 
high-redshift galaxies (e.g., \citealt{rigby08}; \citealt{papovich07}). 
The dust-enshrouded SFRs can be estimated by transforming total infrared 
luminosities ($L_{\rm IR} \equiv L(8-1000~\mu m)$) into SFRs 
\citep{kennicutt98}. 

To convert the 24~$\mu$m emission to a total IR luminosity we followed 
the approach presented in \citet{wuyts08}. Specifically, we used the 
infrared SEDs of star-forming galaxies provided by \citet{dale02}. The 
template set allows us to quantify the IR/MIR flux ratio for different 
heating levels of the interstellar environment, parameterized by 
$dM(U) \sim U^{-\alpha}dU$, where $M(U)$ represents the dust mass heated 
by an intensity $U$ of the interstellar field. We computed the total 
infrared luminosity $L_{\rm IR, \alpha}$ for each object for all \citet{dale02} 
templates within the reasonable range from $\alpha=1$ for active galaxies 
to $\alpha=2.5$ for quiescent galaxies. The mean of the resulting 
$\log{L_{\rm IR,\alpha=1,...,2.5}}$ was adopted as a best estimate for the IR 
luminosity. Table~\ref{tab-3} lists the estimated total IR luminosities, 
$L_{\rm IR}$, with the corresponding 1~$\sigma$ errors, with and without the 
uncertainties due to random photometric redshift errors. The adopted 
approach to estimate $L_{\rm IR}$ from 24~$\mu$m fluxes has been shown to 
produce SFRs that are in better agreement with the SFRs determined from 
SED modeling (\citealt{franx08}; \citealt{wuyts08}) and from dust-corrected 
$H\alpha$ line fluxes \citep{muzzin10}, with respect to the often used 
local luminosity-dependent approaches, which can systematically 
over-estimate SFRs by a factor of 4-6 (\citealt{papovich07}; 
\citealt{murphy09}; \citealt{muzzin10}). More 
importantly, our approach adopted to estimate $L_{\rm IR}$ from 24~$\mu$m 
fluxes is further supported by the detailed analysis of the far-IR SED 
(from 24~$\mu$m to 870~$\mu$m) performed in \citet{muzzin10} on 
two ultra-luminous infrared galaxies (ULIRGs) at $z\sim2$. 

We convert the $L_{\rm IR}$ to SFR using the \citet{kennicutt98} relation, 
$SFR(L_{\rm IR}) = 0.63 \cdot L_{\rm IR} / 5.8\times10^{9}$~L$_{\sun}$, where the 
factor of 0.63 is to convert to a pseudo-\citet{kroupa01}~IMF. The 
estimated SFRs are listed in Table~\ref{tab-3}, along with the 1~$\sigma$ 
errors, with and without the uncertainties due to random photometric 
redshift uncertainties. As shown in Table~\ref{tab-3}, the random 
uncertainties on $L_{\rm IR}$ and SFRs are dominated by the contribution 
from random photometric redshift errors. Systematic template uncertainty 
(not included in the errors in Table~\ref{tab-3}) can contribute additionally 
to the total error budget, with $\pm0.45$~dex variation from 
$\log{L_{\rm IR,\alpha=2.5}}$ to  $\log{L_{\rm IR,\alpha=1}}$ \citep{wuyts08}.

The estimated total IR luminosities of the MIPS detected sources range from 
$\sim5.0\times10^{12}$~L$_{\sun}$ to $\sim4.0\times10^{13}$~L$_{\sun}$, with 
80\% of them having $L_{\rm IR} \geq 10^{13}$~L$_{\sun}$, typical of 
Hyper-Luminous Infrared Galaxies (HLIRGs), and the remaining being ULIRGs. 
The fraction of HLIRGs in the most massive galaxies at 
$3.0 \leq z < 4.0$ is larger by a factor $\sim10$ with respect to the 
fraction of HLIRGs in the \citet{kriek08} sample of massive 
($M_{\rm star} \approx 10^{11}$~M$_{\sun}$) galaxies at $2.0<z<2.7$ 
with spectroscopic redshifts \citep{muzzin10}. Whereas the sample at 
$2.0<z<2.7$ is less massive than our mass-selected sample at 
$3.0 \leq z < 4.0$ by a factor of $\sim3$, the large difference in 
the fraction of HLIRGs seems to suggest a large evolution in the number 
density of massive HLIRGs from $z=3.5$ to $z=2.3$.

As previously noted, the observed 24~$\mu$m emission could arise from 
warm/hot dust and PAH molecules, heated by either dust-enshrouded star 
formation or highly-obscured AGN. 
If all the IR luminosity is associated with dust-enshrouded star formation, 
the resulting SFRs range between $\sim600$ to $\sim4300$~M$_{\sun}$~yr$^{-1}$, 
with the exclusion of C1-22857, for which only an upper limit was derived.  
These values are tens to several hundreds of times larger than the SFRs 
estimated from SED modeling. On average, the SFRs estimated from the 
24~$\mu$m fluxes are a factor of $\sim200$ larger than the SFRs estimated 
from SED modeling. Moreover, three galaxies (C1-6110, C1-15182, and C1-23152) 
have MIPS-derived SFRs of the order of 
1.3-2.5$\times$10$^{3}$~M$_{\sun}$~yr$^{-1}$, whereas the FUV-to-8~$\mu$ observed 
SEDs are consistent with zero star formation. This suggests that if the 
24~$\mu$m flux is from star formation, most of it must be completely 
obscured by dust. The MIPS estimated SFRs translate to specific star formation 
rates $sSFR\approx10^{-8.8}-10^{-7.8}$~yr$^{-1}$, which would imply that the most 
massive galaxies at $z=3.5$ are extremely actively star-forming systems 
that would double their stellar mass in $(0.6-7)\times10^{8}$~yrs if the derived 
SFRs were to be sustained at the current levels. However, little evolution 
seems to have been found in the number density of the most massive galaxies 
from $z=3.5$ to $z=1.6$ \citep{marchesini09}, which would imply a growth in 
stellar mass in the most massive galaxies over this redshift range by 
$\sim$30\% ($>3$ times smaller than the implied growth from $z=3.5$ to 
$z=2.5$ from the MIPS-derived SFRs), although larger evolution would be 
allowed once systematic uncertainties are taken into account. Therefore, 
either the very large star-forming activity indicated by the observed 
24~$\mu$m emission has to be very quickly quenched in the majority of the 
most massive galaxies at $3.0 \leq z < 4.0$, or the MIPS-derived SFRs are 
systematically biased by, e.g., contamination from AGN activity. 

Indeed, in the local universe, AGN is thought to be the dominant source of 
radiation responsible for the far- and mid-IR SEDs of galaxies with 
$L_{\rm IR}\sim10^{13}$~L$_{\sun}$ (e.g., \citealt{genzel98}; \citealt{lutz98}). 
If the MIR emission is dominated by AGN-heated dust, the large fraction 
of very massive galaxies at $3.0 \leq z < 4.0$ with MIPS detection suggests 
that AGNs are very common ($\gtrsim$80\%) in the most massive galaxies at 
these redshifts. While the fraction of AGNs in dropout galaxies at $z>3$ 
is generally estimated to be low ($\sim$3-7\%; \citealt{steidel02}; 
\citealt{laird06}; \citealt{reddy06}), estimates of the AGN fraction in 
massive galaxies at $3.0 \leq z < 4.0$ are still very uncertain. Common 
AGN activity in massive galaxies has been found at lower redshifts, with 
AGN fraction of $\sim30$\% at $z\sim2.5$ (\citealt{papovich06}; 
\citealt{kriek07}). Evidence for an increasing fraction of AGN as a function 
of stellar mass as been also shown by \citet{kriek07}, with AGN fraction 
that could reach as high as 70\% for the most massive galaxies at 
$2.0<z<2.7$. If the observed 24~$\mu$m emission represents a signature of 
AGN-heated dust, then our results represent further supporting evidence 
for higher AGN fractions at high-$z$ and in the most massive galaxies.

With the currently available data it is not possible to discriminate between 
dust-enshrouded starburst or highly-obscured AGN as the dominant source 
responsible in heating the dust in our sample of very massive galaxies. 
Significant contributions from both sources to the observed MIPS~24~$\mu$m 
fluxes cannot be excluded, and their relative importance will certainly 
vary among our sample. However, if most of the massive galaxies in our 
sample have extreme SFRs, as derived from the MIPS data, then it is 
unlikely that we are witnessing short-lived bursts, as the duty cycle of 
the star formation has to be long to account for the observed large fraction 
of MIPS-detected sources. This seems to be in contradiction with the need 
for the extreme MIPS-derived star-formation activity to be rapidly 
($<10^{8}$~years) quenched to avoid overprediction of the high-mass end of 
the SMF of galaxies at $z<3$. Moreover, the estimated $L_{IR}$ are typical 
of HLIRGs, for which AGN is generally thought to be a significant (if not 
dominant) source of radiation responsible for the very large IR 
luminosities. Finally, for the targeted redshift range, the MIPS~24~$\mu$m 
band probes rest-frame wavelengths around $\sim5.5$~$\mu$m, where hot dust 
dominates the MIR emission, and the contribution from the AGN as the source 
of the radiation field heating the dust becomes increasingly more likely. 
For all these reasons, the very high MIPS-estimated star formation rates are 
unlikely, and we therefore favor AGN instead of starburst activity 
as the dominant source of the observe MIPS~24~$\mu$m emission.

Whatever the source of radiation responsible for heating the dust is 
(dust-enshrouded star formation and/or AGN), the very large IR luminosities 
estimated in our sample of very massive galaxies at $3.0 \leq z < 4.0$ show 
that, despite the already very large stellar masses, there is still plenty 
of gas and dust around either the supermassive black holes or the star 
forming regions.

\subsubsection{X-ray emission}

AGNs can be efficiently identified by their X-ray emission, which is 
thought to be due to up-scattered UV photons from the accretion disk. 
AGN-induced X-ray emission can be distinguished from that induced by 
star formation by the hardness ratio and (particularly) the luminosity. 
{\it Chandra} X-ray data are available over both the COSMOS and AEGIS 
fields. We have used the publicly available X-ray catalogs 
(\citealt{laird09} and \citealt{elvis09} for the AEGIS and COSMOS fields, 
respectively) to search for X-ray detections within our mass-selected 
sample at $3.0\leq z < 4.0$.

Three sources, namely C1-15182, C1-19536, and A2-15753, are detected in 
the Chandra images, with hard band (2-7~keV), fluxes of 
(3.8$\pm$1.0)$\times$10$^{-15}$, (1.3$\pm$0.2)$\times$10$^{-14}$, and 
(7.6$\pm$1.2)$\times$10$^{-15}$~erg~s$^{-1}$~cm$^{-2}$, respectively. 
Assuming a power-law photon index $\Gamma=1.9$ \citep{nandra94}, these 
fluxes correspond to X-ray luminosities $L_{\rm 2-7~keV}$ of 
(3.8$\pm$1.3)$\times$10$^{44}$, (1.0$\pm$0.2)$\times$10$^{45}$, and 
(5.7$\pm$1.5)$\times$10$^{44}$~erg~s$^{-1}$, respectively, typical of 
high-luminosity AGNs 
($L_{\rm 2-7~keV}>3\times 10^{43}$~erg~s$^{-1}$).\footnote{The quoted errors 
of $L_{\rm 2-7~keV}$ includes the error due to photometric redshift random 
uncertainties.} Their hardness ratios are 0.17$\pm0.3$, -0.24$\pm0.11$, 
and 0.31$\pm$0.13, typical of narrow-line and high-$z$ obscured AGNs 
\citep{brusa09}. Using Figure~5 in \citet{treister09}, which plots the 
neutral hydrogen column density ($N_{\rm H}$) as a function of hardness 
ratio for 15 high-redshift X-ray sources, we estimated 
$N_{\rm H}=$(4$^{+4}_{-2}$)$\times$10$^{23}$, 
(1$^{+1.0}_{-0.5}$)$\times$10$^{23}$, and (6$\pm$2)$\times$10$^{23}$~cm$^{-2}$, 
characteristic of moderately obscured AGNs. 

We therefore conclude that the X-ray data for the three detected sources 
support the presence of powerful AGNs in all three sources, providing 
further evidence for AGN radiation being the dominant source for heating 
the dust and for the large MIPS fluxes. Note however that only 
high-luminosity AGNs with $L_{\rm 2-7~keV}>10^{43.7-43.9}$~erg~s$^{-1}$ could 
have been detected for sources at $z>3$ given the limiting source detection 
depths of the X-ray data over AEGIS and COSMOS. In other words, the lack 
of X-ray detection does not provide information on the presence, or lack 
thereof, of high-luminosity AGNs in the remaining 11 sources of our 
mass-selected sample.

\subsubsection{Continuum emission from the AGN}

In the previous section, we found that the AGN is likely the dominant 
source of radiation responsible for heating the dust and for the large 
IR luminosities. Moreover, three galaxies have X-ray detections with 
hardness ratios  and estimated X-ray luminosities typical of obscured 
high-luminosity AGNs. Therefore, the AGN emission could potentially 
contribute to the observed SED, biasing the derived stellar masses. 
We investigate this by subtracting the AGN contribution from the observed 
SED and by re-fitting the corrected SED. Specifically, we assume a 
power-law SED for the AGN, with $F_{\nu} \propto \nu^{\alpha}$. The value 
of $\alpha$ has been derived by fitting the rest-frame UV and the MIPS 
24~$\mu$m photometry, with values of $\alpha$ found in the range 
$-2.9<\alpha<-1.7$. The maximum AGN contribution is then set by the 
rest-frame UV fluxes in combination with the 24~$\mu$m band, and subtracted 
from the observed SED. The resulting SEDs are finally re-modeled using FAST 
to derive stellar masses.
We find that the derived stellar masses are smaller by typically 
$\sim$0.08~dex. For two of the three galaxies with X-ray detection, this 
analysis results in stellar masses smaller by only 0.05-0.08~dex, slightly 
larger than their random errors on $M_{\rm star}$. For the third galaxy 
with X-ray detection (A2-15753), the AGN contribution is 0.18~dex, the 
largest in our mass-selected sample (although still much smaller than 
the systematic uncertainties due to different SED-modeling assumptions).

We stress that the estimated systematic effects caused by the AGN 
contributions should be strictly regarded as upper limits, as our 
approach maximizes, by construction, the contribution of the AGN to 
the observed SED. We therefore conclude that these systematic 
effects are in general small, and certainly much smaller than the systematic 
uncertainties caused by the different SED-modeling assumptions and/or by 
potential systematic errors in the photometric redshift estimates (see 
\S~\ref{sec-dustold}). We note that additional contamination of some of 
the medium- and broad-band filter fluxes could be potentially caused by 
the presence of strong AGN line emission.


\section{THE STELLAR MASS FUNCTION AND DENSITY} \label{sec-mf}


\subsection{Methodology}

We used the mass-selected sample defined in \S~\ref{sec-ss} to derive 
more accurate measurements of the high-mass end of the SMF of galaxies 
at $3.0 \leq z < 4.0$. To estimate the observed SMF we have followed 
the analysis in \citet{marchesini09}, which we refer to for a detailed 
descriptions of the methods used in this work. Briefly, we have derived 
the SMF using two methods, the $1/V_{\rm max}$ estimator and a parametric 
maximum likelihood method. 

For the $1/V_{\rm max}$ estimator, we used the extended version as defined 
by \citet{avni80}. The Poisson error in each stellar mass bin was computed 
adopting the recipe of \citet{gehrels86}. As extensively discussed in the 
literature, the $1/V_{\rm max}$ estimator has the advantages of simplicity 
and no a priori assumption of a functional form for the stellar mass 
distribution; it also yields a fully normalized solution. However, it can 
be affected by the presence of clustering in the sample. Field-to-field 
variation represents a significant source of uncertainty in deep surveys, 
since they are characterized by small areas and hence small probed volumes. 
The contribution due to cosmic variance to the total error budget is 
quantified in \S~\ref{sec-errors}.

We also measured the observed SMF using the STY method \citep{sandage79}, 
which is a parametric maximum-likelihood estimator. The STY method has 
been shown to be unbiased with respect to density inhomogeneities (e.g., 
\citealt{efstathiou88}), it has well-defined asymptotic error properties 
(e.g., \citealt{kendall61}) and does not require binning. We have assumed 
that the number density $\Phi(M)$ of galaxies is described by a 
\citet{schechter76} function,

\begin{eqnarray}
\Phi (M) = (\ln{10}) \Phi^{\star} 
\big[ 10^{(M-M^{\star})(1+\alpha)} \big] \times \exp{\big[ -10^{(M-M^{\star})} \big]},
\end{eqnarray}
where $M=\log{(M_{\rm star}/M_{\sun})}$, $\alpha$ is the low mass-end slope, 
$M^{\star}=\log{(M_{\rm star}^{\star}/M_{\sun})}$ is the characteristic stellar 
mass at which the SMF exhibits a rapid change in the slope, and $\Phi^{\star}$ 
is the normalization. Following \citet{marchesini09}, the best-fit solution 
is obtained by maximizing the likelihood $\Lambda$ with respect to the 
parameters $\alpha$ and $M^{\star}$. The value of $\Phi^{\star}$ is then 
obtained by imposing a normalization on the best-fit SMF such that the total 
number of observed galaxies in the sample is reproduced. 


\subsection{Uncertainties on the Stellar Mass Function}\label{sec-errors}

As discussed in \citet{marchesini09}, uncertainties due to small-number 
statistics, photometric redshift errors, cosmic variance, and different 
SED-modeling assumptions contribute to the total error budget of the 
high-mass end of the high-redshift SMF. 

The uncertainties on the SMF due to random photometric redshift errors 
have been estimated following the recipe in 
\citet{marchesini09}.\footnote{In \citet{marchesini09}, systematic 
photometric redshift errors were estimated by adopting different template 
sets to derive the photometric redshifts. Here we decided not to use the 
other template sets distributed with EAZY due to the significantly worse 
resulting $z_{\rm phot}-z_{\rm spec}$ comparisons, whereas in \citet{marchesini09} 
they resulted in $z_{\rm phot}-z_{\rm spec}$ comparison of similar quality, or 
only slightly worse.}
Specifically, for each galaxy in the $K$-selected sample, a set of 200 
mock SEDs was created by perturbing each flux point according to its 
formal error bar. Second, we estimated the photometric redshift in the same 
way as described in \S~\ref{subsec-zphot}. Third, we fit the mock SEDs 
with FAST to estimate stellar masses as described in \S~\ref{subsec-sed}. 
Finally, we have derived SMFs of galaxies with the $1/V_{\rm max}$ and the 
maximum likelihood analysis for each of the 200 Monte Carlo realizations 
of the $K$-selected sample. The contribution to the total error budget 
of the SMF derived using the $1/V_{\rm max}$ method due to random photometric 
redshift errors ($\sigma_{\rm z,ran}$) is listed in Table~\ref{tab-mf1}, and 
is roughly 0.13~dex, about a factor of 1.7 smaller than in 
\citet{marchesini09}. Similarly to what found in \citet{marchesini09}, 
the contribution of random photometric redshift errors on the Schechter 
function parameters of the SMF at $3.0 \leq z < 4.0$ is instead negligible 
with respect to Poisson errors. In fact, Poisson errors largely dominate 
the random error budget of the Schechter function parameters due to the 
complete lack of constraint on the low-mass end slope $\alpha$ (see 
Table~\ref{tab-mf2}). Because the low-mass end slope is not constrained by 
the NMBS dataset, we have also repeated the maximum-likelihood analysis 
fixing the value of the low-mass end slope at $\alpha=-1.0$ (corresponding 
to the value of the low-mass end slope of the SMF of galaxies at 
$1.3 \leq z < 2.0$ from \citealt{marchesini09}) and $\alpha=-1.75$ 
(corresponding to the value of the low-mass end slope of the SMF of 
galaxies at $2.5<z<3.5$ from \citealt{kajisawa09}).

To quantify the uncertainties due to field-to-field variations in the 
determination of the SMF, we proceeded as in \citet{marchesini07a}. 
Briefly, using the $1/V_{\rm max}$ method, we measured $\Phi^{j}$, the galaxy 
number density in the stellar mass bin $\Delta M$ for the $j$th field. 
The contribution to the error budget from cosmic variance is estimated 
using $\sigma_{\rm cv} = rms(\Phi^{j})/\sqrt{2}$. The final 1~$\sigma$ 
random error associated with $\Phi(M)$ is then 
$\sigma=(\sigma_{\rm Poi}^{2}+\sigma_{\rm cv}^{2}+\sigma_{\rm z,ran}^{2})^{1/2}$, 
where $\sigma_{\rm Poi}$ is the Poisson error in each mass bin. The values 
of $\sigma_{\rm Poi}$ and $\sigma_{\rm cv}$ are also listed in 
Table~\ref{tab-mf1}. The contribution to the error budget from cosmic 
variance can also be estimated for a given population using predictions 
from cold dark matter theory and the galaxy bias. We have derived the 
cosmic variance following the cosmic variance cookbook presented by 
\citet{moster10} and using the parameters for our survey and for a 
population of massive galaxies with $M_{\rm star}>10^{11}$~M$_{\sun}$, resulting 
in an uncertainty due to cosmic variance of 0.18~dex, in very good 
agreement with our empirical estimate. 

Whereas we refer to \citet{marchesini09} for a complete analysis and 
discussion of the systematic uncertainties due to different SED-modeling 
assumptions, we have repeated the whole analysis adopting the stellar 
population synthesis models of \citet{maraston05} with a \citet{kroupa01} 
IMF and exponentially declining SFHs with values of the e-folding timescale 
ranging from 100 Myr to 10 Gyr. The resulting systematic uncertainties 
on the SMF measured with the $1/V_{\rm max}$ method are listed in 
Table~\ref{tab-mf1}, and the corresponding values of the Schechter 
function parameters measured with the maximum likelihood analysis are 
listed in Table~\ref{tab-mf2}. The adoption of the \citet{maraston05} 
models result, in general, in smaller derived stellar masses by 
$\sim0.15$~dex. As previously shown in \citet{marchesini09} and 
\citet{muzzin09}, different combinations of adopted metallicity and 
extinction curve also result in systematic effects on the derived stellar 
masses, although to a much smaller extent with respect to the biases 
introduced by different stellar population synthesis models and the 
specific choices of the adopted SFHs. In particular, different assumptions 
on the SFH with respect to those adopted in our work (e.g., two-component 
models of the SFH, or exponentially-increasing SFH) can introduce 
additional systematic biases toward both larger and smaller stellar 
masses (\citealt{wuyts07}; \citealt{lee09}; \citealt{maraston10}).
In \S~\ref{sec-dustold}, we consider the systematic uncertainties due to 
the inclusion of an additional template in the template set used to derive 
photometric redshifts with EAZY. This additional template consists of an 
old (1~Gyr) and dusty ($A_{\rm V}=3$~mag) single stellar population.


\subsection{Stellar Mass Function}

\begin{figure}
\epsscale{1.1}
\plotone{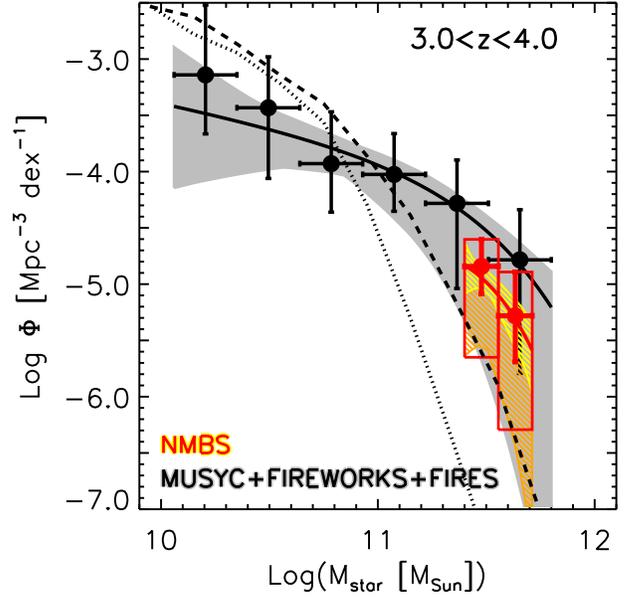}
\caption{SMFs of galaxies at redshift $3.0 \leq z < 4.0$ from the 
NMBS (red, orange, and yellow symbols) and the analysis of 
\citet{marchesini09} (black and gray symbols). The filled symbols 
represent the SMF derived with the $1/V_{\rm max}$ method, with error 
bars showing the total 1~$\sigma$ random errors, including 
photometric redshift errors and field-to-field variations; the red 
boxes include also the systematic uncertainties due to the 
different SED-modeling assumptions adopted (see \S~\ref{subsec-sed}).
The solid curves represent the SMFs derived with the maximum likelihood 
analysis, with shaded regions representing the 1~$\sigma$ uncertainties. 
The black error bars and gray shaded area include the systematic 
uncertainties due to different template sets in the photometric redshift 
estimate. The orange hatched area includes also the systematic 
uncertainties due to the different SED-modeling assumptions adopted in 
our analysis. The dotted and dashed black curves represent the predicted 
SMFs from the semi-analytic model of \citet{somerville08}, where the dashed 
curve is derived from the dotted curve after convolution with a normal 
distribution of standard deviation of 0.25~dex. The NMBS allows us to 
derive more accurate measurements of the high-mass end of the SMF of 
galaxies at $3.0 \leq z < 4.0$. \label{fig-smf}}
\end{figure}

Figure~\ref{fig-smf} shows the SMF of galaxies at redshift 
$3.0 \leq z < 4.0$ derived in this work (colored symbols) compared to the 
SMF of galaxies at $3.0 \leq z < 4.0$ derived in \citet{marchesini09} 
(black and gray symbols). Points with error bars show the SMFs derived 
using the $1/V_{\rm max}$ method. The solid curves show the SMFs derived 
with the maximum likelihood analysis, while the shaded regions represent 
their 1~$\sigma$ uncertainties. The plotted uncertainties of the SMF 
measurements from \citet{marchesini09}, the thick red errors bars, and 
the yellow shaded area represent the total 1~$\sigma$ random errors, 
including cosmic variance and photometric redshift errors as quantified 
in \S~\ref{sec-errors}. The thin red error bars and the orange shaded area 
include also the systematic uncertainties due to the different SED-modeling 
assumptions adopted in this work.

\begin{figure}
\epsscale{1.15}
\plotone{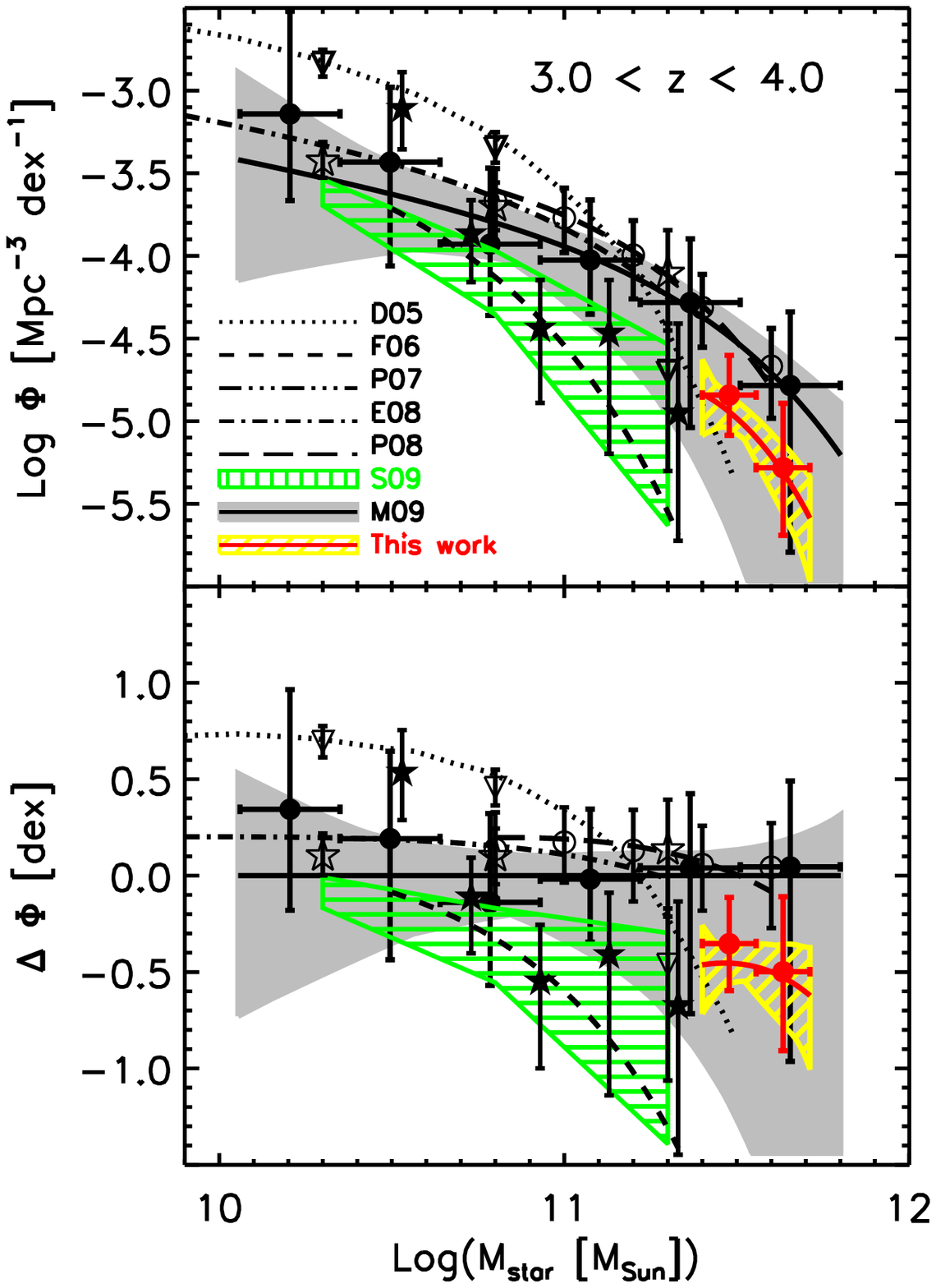}
\caption{{\it Top panel:} Comparison between the SMF at $3.0\leq z < 4.0$ 
(yellow and red) derived from this work and previous measurements from the 
literature, including the measurement from \citet{marchesini09} (gray and 
black). For the SMFs derived from this work and \citet{marchesini09}, the 
filled circles represent the measurements using the $1/V_{\rm max}$ method, 
while the solid curves represent the measurements using the 
maximum-likelihood analysis; the 1~$\sigma$ error bars of the $1/V_{\rm max}$ 
measurements include Poisson errors, field-to-field variations, and 
uncertainties from photometric redshift uncertainties (both random and 
systematic). Similarly for the 1~$\sigma$ error of the maximum-likelihood 
measurements (yellow and gray regions). Previous works are plotted as 
filled stars and dashed curves (\citealt{fontana06}; F06); open circles and 
long-dashed curves (\citealt{perez08}; P08); open stars and dot-dashed curves 
(\citealt{elsner08}; E08); open triangles and dotted curves 
(\citealt{drory05}; D05). The hatched green area shows the SMF of 
$B$-dropout galaxies from \citet{stark09}. {\it Bottom panel:} Symbols as in 
the top panel, but now the differences between the SMFs measurements showed 
in the top panel and the SMF from \citet{marchesini09}, 
$\Delta \Phi = \log{\Phi}-\log{\Phi_{DM09}}$, are plotted as a function of 
stellar mass to highlight the differences. \label{fig-smfcomp}}
\end{figure}

The large surveyed area (i.e., effective area of 0.44~square degrees, 
a factor of $\sim$3 larger than in \citealt{marchesini09}) and the 
accurate photometric redshift estimates allow for the determination of 
the number density of the most massive galaxies at $3.0 \leq z < 4.0$ 
with unprecedented accuracy, as clearly shown by Figure~\ref{fig-smf} 
from the comparison with the SMF previously derived by \citet{marchesini09}.
Figure~\ref{fig-smfcomp} shows the comparison between the SMF derived 
in this work and previous measurements of the SMFs of galaxies at 
$z\sim3.5$. The high-mass end of the SMF measured in our analysis is in 
good agreement with previous measurements.

\begin{figure}
\epsscale{1.1}
\plotone{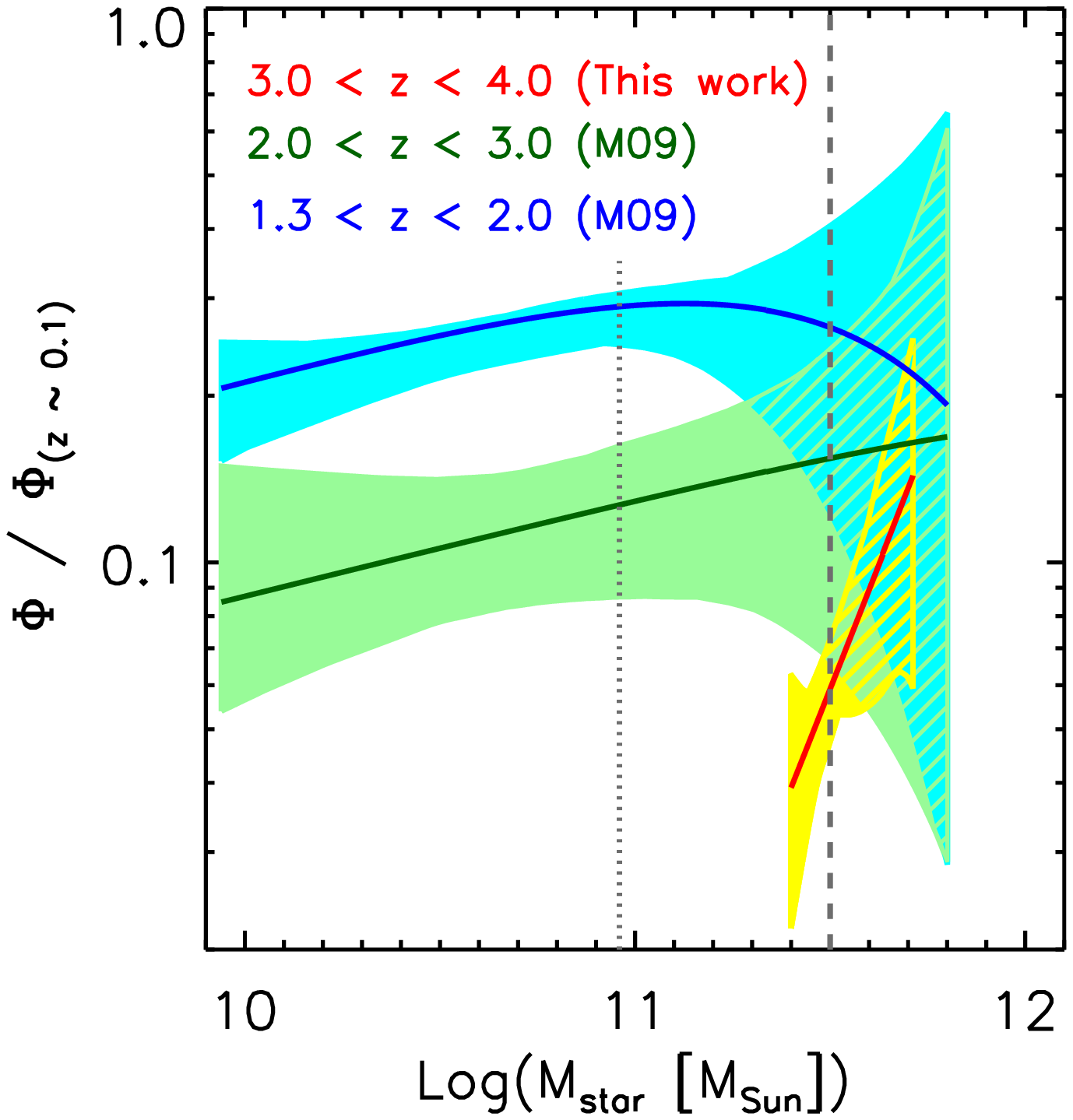}
\caption{Ratio of the high-$z$ SMFs ($\Phi$) and the local SMF 
($\Phi_{z\sim0.1}$; \citealt{cole01}) plotted as function of the stellar 
mass as measured from the maximum-likelihood analysis. The SMFs at 
$z\sim1.6$ (blue) and $z\sim2.5$ (green) are taken from 
\citet{marchesini09}. The shaded regions represent the total 
1~$\sigma$ random uncertainties. The vertical dashed and dotted 
lines represent the value of 3$\times$10$^{11}$~M$_{\sun}$ and the 
$z=0.1$ characteristic stellar mass, $M^{\star}_{\rm star}\sim10^{11}$~M$_{\sun}$, 
respectively. \label{fig-smfevol}}
\end{figure}

\begin{deluxetable}{ccccccc}
\centering
\tablecaption{SMF at $3 \leq z < 4.0$ derived with the $1/V_{\rm max}$ method
\label{tab-mf1}}
\tablehead{\colhead{$\log{M_{\rm star}}$} & \colhead{$\log{\Phi}$} & 
           \colhead{$\sigma$} & 
           \colhead{$\sigma_{\rm Poi}$} &
           \colhead{$\sigma_{\rm z,ran}$} & 
           \colhead{$\sigma_{\rm cv}$} &
           \colhead{$\sigma_{\rm MA05,sys}$} \\
                    ($M_{\sun}$) & (Mpc$^{-3}$~dex$^{-1}$) &
                     & & & & }
\startdata
11.63 & -5.282 & $^{+0.390}_{-0.409}$ & $^{+0.253}_{-0.282}$ & 0.151 & 0.256 & $^{+0.0}_{-0.6}$ \\
11.48 & -4.842 & $^{+0.240}_{-0.244}$ & $^{+0.146}_{-0.153}$ & 0.117 & 0.150 & $^{+0.0}_{-0.6}$ \\
\enddata
\tablecomments{$\sigma=(\sigma^{2}_{Poi}+\sigma^{2}_{cv}+\sigma^{2}_{z,ran})^{1/2}$ 
is the total 1~$\sigma$ random error, including the Poisson errors 
($\sigma_{Poi}$), the errors due to random photometric redshift  
uncertainties ($\sigma_{z,ran}$), and the error due to cosmic variance 
($\sigma_{cv}$; see \S~\ref{sec-errors}); $\sigma_{\rm MA05,sys}$ is the 
systematic uncertainty due to the different SED-modeling assumptions, 
i.e., the \citet{maraston05} stellar population synthesis 
models, \citet{kroupa01} IMF, solar metallicity, 
\citet{calzetti00} extinction curve, and exponentially 
declining SFHs (see \S~\ref{subsec-sed} and \S~\ref{sec-errors}).}
\end{deluxetable}

\begin{deluxetable}{ccc}
\centering
\tablecaption{Best-fit Schechter function parameters of the SMF
\label{tab-mf2}}
\tablehead{\colhead{$\alpha$}     & 
           \colhead{$\log{M^{\star}_{\rm star}}$}      & 
           \colhead{$\Phi^{\star}$}          \\
                     &
                    ($M_{\sun}$) & (10$^{-5}$~Mpc$^{-3}$~dex$^{-1}$)}
\startdata
                     $0.6^{+4.4}_{-5.4}~(2.0^{+4.0}_{-6.0})$ & 
                     $10.97^{+1.54}_{-0.38}~(10.77^{+1.23}_{-0.34})$ & 
                     $2.00^{+14.93}_{-1.90}~(0.24^{+113.48}_{-0.23})$\\
                     $-1.0~(-1.0)$ & 
                     $11.17^{+0.18}_{-0.14}~(11.05^{+0.31}_{-0.22})$ & 
                     $3.73^{+5.35}_{-2.10}~(2.58^{+14.2}_{-2.06})$\\
                     $-1.75~(-1.75)$ & 
                     $11.30^{+0.28}_{-0.18}~(11.15^{+0.45}_{-0.25})$ & 
                     $2.90^{+6.37}_{-2.13}~(2.62^{+20.12}_{-2.38})$\\
\enddata
\tablecomments{The quoted error correspond to the 
1~$\sigma$ error estimated from the maximum-likelihood 
analysis as described in \S~\ref{sec-errors}. The values 
in parenthesis are the best-fit Schechter function 
parameters derived with different SED-modeling assumptions, 
i.e., the \citet{maraston05} stellar population synthesis 
models, \citet{kroupa01} IMF, solar metallicity, 
\citet{calzetti00} extinction curve, and exponentially 
declining SFHs (see \S~\ref{subsec-sed} and 
\S~\ref{sec-errors}). Note that the low-mass end slope is 
completely unconstrained. The second and third rows list 
the best-fit Schechter function parameters obtained with 
fixed $\alpha$.}
\end{deluxetable}

Combined with the results from \citet{marchesini09}, the number density of 
the most massive galaxies appears to have evolved by a factor of $\sim2$ 
from $z=3.5$ to $z=2.5$, and by a factor of $\sim3$ from $z=3.5$ to $z=1.6$.
This is shown in Figure~\ref{fig-smfevol}, where the ratio of the high-$z$ 
SMFs and the local SMF from \citet{cole01} is plotted as a function of the 
stellar mass. However, due to the steepness of the high-mass end, the 
implied evolution of the number density translates to small growth in 
stellar mass of the most massive galaxies, by 30\%-40\% from $z=3.5$ to 
$z=1.6$, and by $\sim$40\% from $z=1.6$ to $z=0.1$, although systematic 
uncertainties allow for a larger evolution. Recently, \citet{vandokkum10} 
have estimated a growth of a factor of $\sim2$ in the stellar mass of 
massive galaxies from $z=2$ to $z=0.1$, in apparent contradiction with our 
results. However, the selection of the sample of massive galaxies in 
\citet{vandokkum10} was very different than ours, as galaxies were selected 
at a constant number density of $n=2\times10^{-4}$~Mpc$^{-3}$ over the 
targeted redshift range. This approach selects galaxies with stellar masses 
$\log{M_{\rm star}}=11.45\pm0.15$ at $z=0.1$, and $\log{M_{\rm star}}=11.15\pm0.15$ 
at $z=2$, a factor of $\sim1.2$ and $\sim2.1$ smaller than the typical 
galaxy in our mass-selected sample at $z=3.5$. As a consequence of the 
mass-dependent evolution derived in \citet{marchesini09}, a smaller growth 
in the stellar mass of massive galaxies would therefore be derived if the 
selection were done at a value of the number density typical of galaxies 
with stellar masses $\log{M_{\rm star}}=11.5$ at $z=0.1$ (the average mass of 
our mass-selected sample at $z=3.5$). We also note that systematic 
uncertainties in the SED modeling, as well as the choice of the $z\sim0$ 
benchmark, can play an important role in the estimate of the evolution of 
the stellar content in massive galaxies. Therefore, we conclude that the 
results in \citet{vandokkum10} are broadly consistent with ours once the 
different selection and systematic uncertainties are taken into account.

Figure~\ref{fig-smf} also shows the SMF predicted from the semi-analytic 
model of \citet{somerville08} (dotted black curve)\footnote{In the 
comparison with the model predictions, we decided to use only the 
model of \citet{somerville08} as it was the model showing the smallest 
disagreements at the high-mass end in the comparison presented in 
\citet{marchesini09}}. This model, built on 
the previous models described in \citet{somerville99} and 
\citet{somerville01}, presents several improvements, including, but not 
limited to, tracking of a diffuse stellar halo component built up of 
tidally destroyed satellites and stars scattered in mergers, galaxy-scale 
AGN-driven winds, fueling of black holes with hot gas via Bondi accretion, 
and heating by radio jets. The prediction from the \citet{somerville08} 
semi-analytic model are taken from their fiducial WMAP-3 model, which 
adopts a fraction $f_{\rm scatter}=0.4$ of the stars in merged satellite 
galaxies added to a diffuse component distributed in a very extended halo 
or envelope. The dashed curve in Figure~\ref{fig-smf} represents the 
model-predicted SMF convolved with a normal distribution of standard 
deviation 0.25~dex, intended to represent measurement errors in 
$\log{M_{\rm star}}$ \citep{fontanot09}\footnote{We note that the typical 
random error on the stellar masses for the galaxies in our 
$3.0 \leq z < 4.0$ sample is smaller than 0.25~dex by a factor of $\sim2$, 
due to the combination of accurate photometric redshift estimates and 
well-sampled SEDs delivered by the NMBS}.

The comparison between the model-predicted and the NMBS-derived SMF 
provides further supporting evidence for the deficit of very massive 
galaxies at $3.0 \leq z < 4.0$ in the theoretical models of galaxy 
formation, a disagreement that was only marginal with the previously 
derived SMFs. Without the inclusion of the systematic uncertainties, 
the disagreement between the observed and the (convolved) model-predicted 
high-mass end of the SMF of galaxies at $3.0 \leq z < 4.0$ is significant 
at the 3~$\sigma$ level. The significance of the disagreement is reduced 
to only 1~$\sigma$ if we include the systematic uncertainties due to 
different SED-modeling assumptions as estimated in \S~\ref{sec-errors} 
(i.e., adopting the \citet{maraston05} instead of the \citet{bruzual03} 
models, and different SFHs). We note that systematic uncertainties due to 
an evolving IMF can play an additional role in reducing the disagreement 
between observed and model-predicted SMFs.

\subsection{Number and Stellar Mass Densities}

The number density, $\eta$, and stellar mass density, $\rho$, in massive 
galaxies at $3.0 \leq z < 4.0$ has been estimated by integrating the SMF 
at $M_{\rm star} > 10^{11.40}$~M$_{\sun}$.\footnote{The SMF has been integrated 
using $M_{\rm star}=10^{13}$~M$_{\sun}$ as the upper limit of the integral. Due 
to the exponential behavior of the SMF at the high-mass end, the estimated 
stellar mass density does not depend significantly on the specific value 
of this limit.}  These values are listed in 
Table~\ref{tab-dens} together with the corresponding total 1~$\sigma$ 
errors. Table~\ref{tab-dens} also lists the 1~$\sigma$ lower limits of 
the number and stellar mass densities of galaxies more massive than 
$10^{8}$~M$_{\sun}$, as well as the densities estimated with the second 
set of SED-modeling assumptions adopted (see \S~\ref{subsec-sed}). The 
number and stellar mass densities estimated by fixing the low-mass end 
slope at $\alpha=-1.0$ and $\alpha=-1.75$ are also listed in 
Table~\ref{tab-dens}.
Compared to the total stellar mass density in galaxies with 
$M_{\rm star}>10^{10}$~M$_{\sun}$ at $3.0 \leq z < 4.0$ estimated by 
\citet{marchesini09}, the contribution of the galaxies in our mass 
selected-sample is $\sim8^{+13}_{-3}$\% (not including systematic errors). 
Note however that this estimate depends strongly on the value of the 
low-mass end slope of the SMF derived in \citet{marchesini09}, which is 
still very poorly constrained (e.g., \citealt{kajisawa09}; 
\citealt{marchesini09}; \citealt{reddy09}).

\begin{deluxetable}{lcc}
\centering
\tablecaption{Number and stellar mass densities at $3 \leq z < 4.0$
\label{tab-dens}}
\tablehead{\colhead{} & \colhead{$M_{\rm star}>10^{11.40}$~M$_{\sun}$} & 
                        \colhead{$M_{\rm star}> 10^{8.0}$~M$_{\sun}$} }
\startdata
$\log{(\eta}$~[Mpc$^{-3}$]$)$          & -5.55$^{+0.19}_{-0.17}$~(-6.04$^{+0.20}_{-0.18}$) & $>-5.47$~($>-6.00$) \\
$\log{(\rho}$~[M$_{\sun}$~Mpc$^{-3}$]$)$ &  6.00$^{+0.30}_{-0.10}$~(5.49$^{+0.36}_{-0.13}$) & $>~6.09$~($>~5.59$) \\
 & \\
\hline
 & \\
$\log{(\eta}$~[Mpc$^{-3}$]$)$          & -5.55$^{+0.17}_{-0.17}$~(-6.04$^{+0.19}_{-0.17}$) & -3.61$^{+0.71}_{-0.65}$~(-3.79$^{+1.33}_{-1.03}$) \\
$\log{(\rho}$~[M$_{\sun}$~Mpc$^{-3}$]$)$ &  6.00$^{+0.12}_{-0.10}$~(5.48$^{+0.15}_{-0.11}$) & 6.74$^{+0.31}_{-0.26}$~(6.45$^{+0.62}_{-0.42}$) \\
 & \\
\hline
 & \\
$\log{(\eta}$~[Mpc$^{-3}$]$)$          & -5.55$^{+0.18}_{-0.17}$~(-6.05$^{+0.20}_{-0.17}$) & -1.90$^{+0.71}_{-0.69}$~(-2.06$^{+1.28}_{-1.03}$) \\
$\log{(\rho}$~[M$_{\sun}$~Mpc$^{-3}$]$)$ &  6.01$^{+0.14}_{-0.10}$~(5.48$^{+0.18}_{-0.10}$) & 7.25$^{+0.37}_{-0.34}$~(7.04$^{+0.70}_{-0.59}$) \\
\enddata
\tablecomments{Number density, $\eta$, and stellar mass density, 
$\rho$, at $3.0 \leq z < 4.0$ estimated by integrating the best-fit 
Schechter SMF over the specified stellar mass range. The quoted 
1~$\sigma$ errors include Poisson errors, errors due to photometric 
redshift uncertainties, and errors due to cosmic variance. The values 
in parenthesis are the results corresponding to the different 
SED-modeling assumptions, i.e., the \citet{maraston05} stellar 
population synthesis models, \citet{kroupa01} IMF, solar metallicity, 
\citet{calzetti00} extinction curve, and exponentially declining 
SFHs (see \S~\ref{subsec-sed} and \S~\ref{sec-errors}). The third 
and fifth rows list the number densities estimated fixing the low-mass 
end slope at $\alpha=-1.0$ and $\alpha=-1.75$, respectively; the fourth 
and sixth rows list the stellar mass densities estimated fixing the 
low-mass end slope at $\alpha=-1.0$ and $\alpha=-1.75$, respectively.}
\end{deluxetable}

In recent years, there have been several claims of the existence of a 
population of massive (and evolved) galaxies at even larger redshifts, 
i.e., $z\gtrsim4-5$ (e.g., \citealt{yan06}; \citealt{wiklind08}; 
\citealt{mancini09}). 

In particular, \citet{wiklind08} reported of a significant population 
of massive galaxies at $4.9\leq z < 6.5$ found in the 125~arcmin$^{2}$ 
GOODS-South field dominated by old stellar populations with 
$M_{\rm star}=(0.3-3)\times10^{11}$~M$_{\sun}$. In their sample there are 
only two object with $M_{\rm star}>2.5\times10^{11}$~M$_{\sun}$ (one already 
identified by \citealt{mobasher05} as a candidate for a massive, evolved 
galaxy at $z\sim6.5$), implying a stellar mass density at $z\sim5.7$ of 
$\rho(M_{\rm star}>10^{11.40}~M_{\sun})\approx 2\times10^{6}$~M$_{\sun}$~Mpc$^{-3}$, 
and suggesting no evolution of the stellar mass density in the most massive 
galaxies over the $\sim800$~Myr interval from $z\sim5.7$ to $z=3.5$. 
However, \citet{dunlop07} have concluded that there is no convincing 
evidence for any galaxy with stellar mass $M_{\rm star}>2\times10^{11}$~M$_{\sun}$ 
and $z>4$ in the GOODS-South field, which implies a much stronger evolution 
of the stellar mass density in very massive galaxies in the first 1.5~Gyr 
of the universe. \citet{wiklind08} also estimated a stellar mass density 
in galaxies with $M_{\rm star}>10^{10.8}~M_{\sun}$ of 
$\rho > 5\times10^{6}$~M$_{\sun}$~Mpc$^{-3}$ (after correction for the different 
IMF). Combining the results from our analysis and \citet{marchesini09}, it 
implies an increase of the stellar mass density in massive galaxies by a 
factor of $\sim1.6$ from $z\sim5.7$ to $z=3.5$. 

SMFs and stellar mass densities have also been estimated for $V$- 
and $i$-dropout galaxies at $z\sim5$, and $z\sim6$, respectively 
(\citealt{yan06}; \citealt{eyles07}; \citealt{stark07}; \citealt{mclure09}; 
\citealt{stark09}). A comparison with these results is not straightforward. 
First, these are all optically-selected samples, which can be potentially 
biased against massive and evolved galaxies. In contrast, our sample is a 
mass-complete sample constructed from a $K$-selected catalog. Second, the 
stellar mass ranges probed by these studies are very different from ours. 
Our mass-selected sample probes the most massive galaxies, i.e., those with 
$M_{\rm star}>2.5\times10^{11}$~M$_{\sun}$, whereas all the above works probe 
galaxies with typically much smaller stellar masses, i.e., in the range 
$10^{9}-10^{11}$~M$_{\sun}$. A rough estimate of the evolution of the stellar 
mass density in galaxies more massive than $2\times10^{9}$~M$_{\sun}$ can be 
performed by comparing the stellar mass densities derived by the above 
studies with the stellar mass density obtained by combining our results 
with those from \citet{marchesini09} at lower stellar masses, and 
extrapolating the Schechter function to the stellar mass limit probed 
by the above studies. After correcting for the different IMFs, the stellar 
mass density in galaxies more massive than $2\times10^{9}$~M$_{\sun}$ at 
$z\sim5$ and $z\sim6$ is $\rho=(2.3-6.3) \times 10^{6}$~M$_{\sun}$~Mpc$^{-3}$ 
(\citealt{mclure09}; \citealt{stark07}; \citealt{stark09}), and 
$\rho=(0.7-4.3) \times 10^{6}$~M$_{\sun}$~Mpc$^{-3}$ (\citealt{yan06}; 
\citealt{eyles07}; \citealt{mclure09}; \citealt{stark09}), respectively. 
At $z=3.5$, we estimate $\rho=1.5 \times 10^{7}$~M$_{\sun}$~Mpc$^{-3}$, which 
implies an evolution of the stellar mass density in galaxies with 
$M_{\rm star}>2\times10^{9}$~M$_{\sun}$ by a factor of $\sim2-7$ and $\sim3-22$ 
from $z\sim5$ and $z\sim6$ to $z=3.5$, respectively. 
We stress however that the estimated evolution from $z>4$ is very uncertain, 
and affected by both large uncertainties in the SMF of galaxies at $z>4$ 
(especially at the high-mass end), as well as poor constraints on the 
low-mass end slope of the SMF of galaxies at $3.0 \leq z < 4.0$.


\section{MASSIVE, OLD AND DUSTY GALAXIES AT $2<Z<3$?} \label{sec-dustold}

In this section we consider the case of adding an additional template 
to the EAZY template set used to derive photometric redshifts.

Previous searches for old and massive galaxies at $z>4$ highlighted the 
difficulty in unambiguously identifying old and massive objects at extreme 
redshifts on the basis of spectral fitting. In particular, \citet{dunlop07} 
have shown that equally acceptable solutions could be obtained at $z\sim5$ 
with high stellar masses ($M_{\rm star}\sim$6$\times$10$^{11}$~M$_{\sun}$) and 
low extinction ($A_{\rm V}\sim0.4$~mag) and at $z\sim2$ with moderate 
stellar masses ($M_{\rm star}\sim$7$\times$10$^{10}$~M$_{\sun}$) and high 
extinction ($A_{\rm V}\sim3.8$~mag). 

To robustly estimate photometric redshifts, the template set needs to be 
large enough that it spans the broad range of multi-band galaxy colors and 
small enough that the color and redshift degeneracies are kept to a minimum 
(e.g., \citealt{benitez00}). The default template set used in this work was 
carefully constructed and tested in \citet{brammer08}. It has been shown to 
satisfy the requirements for a satisfactory template set, providing 
significantly reduced systematic effects and smaller scatter in the 
$z_{\rm phot}$ versus $z_{\rm spec}$ at all redshifts. This template set 
already includes a dusty starburst model (50~Myr old and $A_{\rm V}=2.75$~mag). 
Here we include an additional template representative of an old (1~Gyr; 
$\tau=100$~Myr) and very dusty ($A_{\rm V}=3$~mag) galaxy, similar to the 
reddest template used in \citet{blanton07}, and we repeat the whole 
analysis. We note that in the local Universe old stellar populations are 
usually not (very) dust-obscured, and that it remains to be seen whether 
this template is physically plausible for the high-redshift Universe.

For eight galaxies (C1-4890, C1-6110, C1-7340, C1-15367, C1-18825, A2-6835, 
A2-15753, and A2-18070), the resulting photometric redshift estimates 
formally lie at $z<3$. The three objects in the AEGIS field have now 
$z\sim2.9$, hence just below our redshift selection window, consistent 
with their redshift probability functions and with A2-15753 having 
rest-frame UV color typical of an LBGs. The five objects in the COSMOS 
field are instead shifted to much lower redshifts, i.e., $z\sim2.4$, 
significantly lower than allowed for by their redshift probability 
functions plotted in Figure~\ref{fig-eazy1}. 

We used FAST to refit stellar population synthesis models to the eight 
galaxies that moved to $z<3$. Six of the eight are best fitted by an old 
stellar population (age$\approx$2~Gyr, i.e., as old as the age of the 
universe at their redshifts), large stellar masses 
($M_{\rm star}\approx 2 \times$10$^{11}$~M$_{\sun}$), and large values of 
extinction ($A_{\rm V}\approx2.1$~mag). The remaining two objects are instead 
best fitted by very dusty ($A_{\rm V}\approx3.2$~mag), young 
(age$\approx$50~Myr) starbursts. 

\begin{figure}
\epsscale{1.1}
\plotone{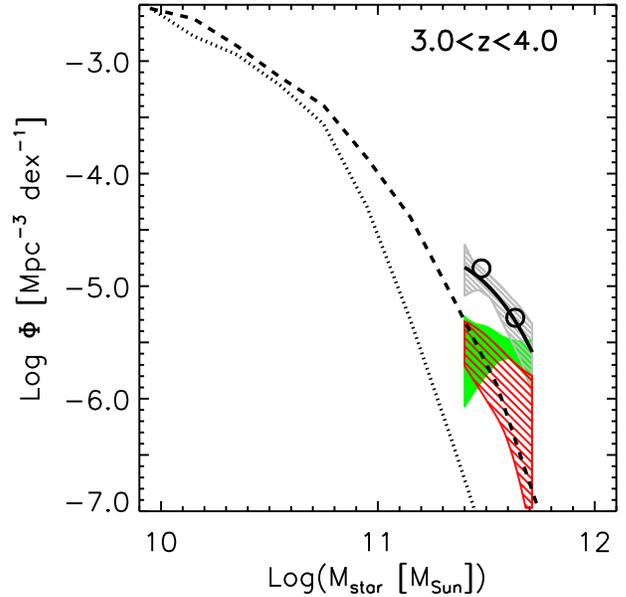}
\caption{SMFs of galaxies at $3 \leq z < 4.0$ measured by adopting 
the additional old and dusty template in the estimate of the 
photometric redshifts. The gray hatched region, and the black 
solid line and empty circles represent the SMF measured adopting 
the default EAZY template set. The colored regions represent the 
SMFs derived by adopting the additional old and dusty template 
in the estimate of the photometric redshifts. The green region 
is obtained by using the \citet{bruzual03} stellar population 
synthesis models, while the hatched red region is obtained by using 
the \citet{maraston05} models. The dotted and dashed curves 
represent the predicted SMFs from the semi-analytic model of 
\citet{somerville08}. \label{fig-smfsys}}
\end{figure}

The SMFs of galaxies at $3.0 \leq z < 4.0$ measured by adopting the 
additional old and dusty template in the estimate of the photometric 
redshifts are shown in Figure~\ref{fig-smfsys}. As shown in this figure, 
the systematic effect on the derived SMF of galaxies at $3.0 \leq z < 4.0$ 
due to the inclusion of this additional (although generally not included) 
template is significant, larger than the systematic effect caused by the 
stellar population synthesis model assumption alone, and bringing the 
observed SMF of galaxies at $3.0 \leq z < 4.0$ in better agreement with 
the SMF predicted from semi-analytic models. 

We note however that the EAZY and FAST best-fit models using the additional 
old and dusty template do not provide statistically better modeling of the 
observed SEDs with respect to the default template set without the old and 
dusty template. This further shows the ambiguity and difficulty in 
characterizing the population of massive galaxies at $z>3$, even with our 
adopted dataset, which is characterized by exquisite wavelength coverage 
from the ultra-violet to the mid-infrared.

In summary, our sample of massive galaxies at $3 \leq z<4$ could potentially
be contaminated (up to $\sim$50\%) by a previously unrecognized population of
massive, old, and very dusty galaxies at $z\sim2.6$. We note that the 
existence of such a population would be an important and puzzling result 
in itself.


\section{SUMMARY AND CONCLUSIONS} \label{sec-disc}

In this paper we have used the far-ultraviolet to mid-infrared coverage 
of the NMBS to derive the observed and rest-frame properties of a complete 
sample of galaxies at $3.0 \leq z < 4.0$ with 
$M_{\rm star} > 2.5 \times 10^{11}$~M$_{\sun}$, and to provide more accurate 
measurements of the high-mass end of the SMF of galaxies at 
$3.0 \leq z < 4.0$. With the addition of five medium-bandwidth NIR filters, 
NMBS delivers accurate photometric redshift, 
$\sigma_{\rm z}/(1+z)\approx 0.02$, for $K$-selected sources at $z>1.5$, 
and provides well-sampled SEDs in the critical wavelength regime around 
the Balmer/4000~\AA~ breaks, allowing, for the first time, the accurate 
detection of the Lyman and the Balmer/4000~\AA~ breaks simultaneously. 
Combined with its large surveyed area, $\sim0.5$~squared degree, it allowed 
us to construct a statistical significant and representative sample of 14 
very massive ($M_{\rm star}>2.5\times 10^{11}$~M$_{\sun}$) galaxies over the 
redshift range $3.0 \leq z < 4.0$.

The typical very massive galaxy at $3.0 \leq z < 4.0$ is red and 
faint in the observer's optical, with a median $r$-band magnitude of 
$\langle r_{\rm tot}\rangle=26.1$. The median $H-K$ color is 1.2, with 
only two object having $H-K<0.9$ at a significant level, highlighting the 
efficiency of the $H-K$ color technique in selecting galaxies at $z>3$ 
with prominent breaks in the rest-frame optical \citep{brammer07}. The 
median rest-frame $U-V$ color $\langle U-V \rangle=1.6$ is similar to local 
Sb spiral galaxies, although we find a range in $U-V$ colors, from 
the typical color of nearby irregular to those of local elliptical galaxies.
The median $U-V$ slope is $\langle \beta \rangle=-0.36$, indicating a 
relatively flat spectrum in $F_{\rm \lambda}$. Intriguingly enough, the 
distribution of UV slopes of the mass-selected sample at $3.0 \leq z < 4.0$ 
is very different from the distributions of UV slopes of UV-selected 
galaxies at $z>2.5$ as well as the $H-K$ galaxies at $z\sim3.7$ so far 
discovered, which 
show distributions of $\beta$ peaked at $\beta < -1.6$. This difference is 
most likely due to the very different ranges in stellar mass probed by the 
different samples with the $H-K>0.9$ galaxies in \citet{brammer07} being 
on average a factor of $\sim$15 less massive than our sample, and the 
typical UV-selected galaxies having masses in the range 
$10^{9}-10^{11}$~M$_{\sun}$ (e.g, \citealt{shapley01}; \citealt{magdis10}). 

By constructing a mass-limited sample from a $K$-selected catalog with 
accurate photometric redshifts rather than the typical color-selection 
techniques, we were able to find a population of galaxies mostly 
complementary to the typical population of dropout galaxies at $z\sim3-4$. 
Specifically, we have shown that only 57\% of the mass-selected sample 
have observed optical colors that satisfy either the $U$- or the $B$-dropout 
color criteria. However, $\sim$50\% of these galaxies are too faint in the 
observed optical to be included in typical spectroscopic samples of LBGs. 

From the SED modeling, our complete sample of massive galaxies at 
$3.0 \leq z < 4.0$ seems to show a range in stellar population properties. 
About 40\%-60\% of the sample is characterized by ages consistent with 
the age of the universe at the targeted redshifts, suggesting that 
the bulk of the stellar mass in these systems was formed at very early 
times. Dust seems to be quite ubiquitous in massive galaxies at 
$3.0 \leq z < 4.0$, with a median extinction of 
$\langle A_{\rm V} \rangle=1.0$~mag. About 30\% of the sample have SFR 
estimates from SED modeling consistent with no star formation activity, 
while the rest of the sample is characterized by significant star 
formation activity, as high as several hundreds solar masses per year. 
Of particular interest is the $z=3.54$ galaxy C1-22857, which has an 
estimated stellar mass of $\sim3\times10^{11}$~M$_{\sun}$, a maximally-old 
age, and completely suppressed star formation. This galaxy is also 
not detected in the {\it Spitzer}-MIPS 24~$\mu$m data, further supporting 
its quiescent nature. Recent ultra-deep NIR spectroscopic observations 
have confirmed a massive galaxy at $z=2.18$ with strongly suppressed star 
formation, pushing back its formation redshift to $z\gtrsim4-7$ 
\citep{kriek09a}. Spectroscopic confirmation of the quiescent nature of 
C1-22857, as well as other objects in our mass-selected sample, is of 
paramount importance, as it would provide even stronger evidence that 
massive galaxies formed their stars extremely efficiently very early 
in time.

Surprisingly, most ($>80$\%) of the massive galaxies $3.0 \leq z < 4.0$ are 
detected in the MIPS~24~$\mu$m data. The total IR luminosities estimated 
from the observed 24~$\mu$m fluxes range from $5\times10^{12}$ to 
$4.0\times10^{13}$~L$_{\sun}$, typical of ULIRGs and HLIRGs, implying extreme 
dust-enshrouded star-formation rates ($\sim$600-4300~M$_{\sun}$~yr$^{-1}$, 
tens to several hundreds of times larger than the SFRs estimated from SED 
modeling), or very common heavily-obscured AGNs, or both in the most 
massive galaxies at $z=3.5$. Whereas it is not possible to discriminated 
between AGN or starburst as the dominant source responsible in heating the 
dust, we favor AGN as a significant, if not a dominant contributor. 
Specifically, the reasons for this are trifold. First, the extreme 
MIPS-derived SFRs cannot be sustained for more than $\sim10^{8}$~yrs 
without overprediction of the high-mass end of the SMFs of galaxies at $z<3$. 
This seems in contradiction with the very large fraction of MIPS detections, 
which implies long duty cycle of the star formation. Second, 80\% of the 
MIPS-detected sources are HLIRGs. In the local universe, AGN is thought to 
be the dominant source of radiation responsible for the far- and mid-IR 
SEDs of HLIRGs. Third, for the targeted redshift range, the 24~$\mu$m 
band probes the rest-frame wavelengths from $\sim4.8$~$\mu$m to 
$\sim7.1$~$\mu$m, where hot dust dominates the MIR emission, and the 
contribution from an AGN as the source of the radiation field heating the 
dust becomes increasingly more likely. If the MIPS-24~$\mu$m emission is 
dominated by AGN-heated dust, the large fraction of very massive galaxies 
at $3.0 \leq z < 4.0$ with MIPS detection suggests that AGNs are very common,
providing further supporting evidence for the coevolution of massive 
galaxies and AGN. We note that three galaxies are detected in the X-ray, 
with 2-7~keV luminosities and hardness ratios typical of obscured, 
high-luminosity AGNs. Observations at longer wavelengths (e.g., in the 
far-IR), as well as other signatures of AGN (e.g., detection of 
narrow-emission lines) are necessary to constrain the occurrence of AGNs 
in this sample and to discriminate between dust-enshrouded star formation 
and heavily-obscured AGN.

We have significantly improved the measurements of the high-mass end of 
the SMF of galaxies at $3.0 \leq z < 4.0$ (the complete analysis of the 
evolution of the SMF of galaxies over the redshift range 
$0.5 \lesssim z < 4.0$ from NMBS will be presented in Marchesini et al., 
in prep.). The accurate photometric redshifts and the large surveyed 
area allowed us to significantly reduce the contributions  of photometric 
redshift errors and cosmic variance to the total error budget. The 
measured high-mass end is in very good agreement with previous measurements, 
providing further supporting evidence for the existence of a 
significant number of very massive galaxies at $z=3.5$. Combined with the 
results from \citet{marchesini09}, the number density of the most massive 
galaxies appears to have evolved little from $z=3.5$ to $z=1.6$, with 
a larger subsequent evolution down to $z\sim0.1$. These results are broadly 
consistent with the growth of a factor of $\sim2$ in the stellar mass of 
massive galaxies from $z=2$ to $z=0.1$ recently estimated in 
\citet{vandokkum10}, once the different sample selection and systematic 
uncertainties are taken into account. The contribution of 
$M_{\rm star}> 2.5\times10^{11}$~M$_{\sun}$ galaxies to the total stellar mass 
budget at $3.0 \leq z < 4.0$ in galaxies with $M_{\rm star}>10^{10}$~M$_{\sun}$ 
is $\sim8^{+13}_{-3}$\%, although this estimate strongly depends on the value 
of the low-mass end slope of the SMF, which is still very poorly constrained. 
The stellar mass density in galaxies more massive than 
$2\times10^{9}$~M$_{\sun}$ seems to evolve by a factor of $5\pm3$ and 
$13\pm10$ from $z\sim5$ and $z\sim6$, respectively, to $z=3.5$.

Our measurement of the high-mass end of $3.0 \leq z < 4.0$ seems to 
exacerbate the disagreement between the observed number densities of 
massive galaxies and those predicted by the latest generation of galaxy 
formation models (e.g., \citealt{somerville08}). The disagreement between 
the observed and the model-predicted high-mass end of the SMF at 
$3.0 \leq z < 4.0$ is significant at the $\sim3$~$\sigma$ level if only 
random errors are considered. However, systematic errors dominate now the 
total error budget at $3.0 \leq z < 4.0$, leading to uncertainties of a 
factor of $\sim8$ in the densities at the high-mass end. When systematic 
uncertainties due to different SED-modeling assumptions are included, the 
found disagreement between observed and model-predicted SMFs is only 
marginally significant. We finally note that additional systematic 
uncertainties on the high-mass end of the $3.0 \leq z < 4.0$ SMF could be 
potentially introduced by either 1) the intense star-formation activity 
and/or the very common AGN activity as inferred from the MIPS~24~$\mu$m 
detections, and/or 2) contamination by a significant population of massive, 
old, and dusty galaxies at $z\sim2.6$ previously unrecognized. This might 
indicate that the high-mass end of the SMF cannot be properly constrained 
without further spectroscopic data.

The NMBS has allowed us to study with unprecedented accuracy the population 
of very massive galaxies at $3.0 \leq z < 4.0$, thanks to its wide surveyed 
area, the accurate photometric redshifts, and the well-sampled SEDs in the 
rest-frame optical. To further improve the characterization of the galaxy 
population at the high-mass end at $3.0 \leq z < 4.0$, it is necessary to 
significantly increase the sample size, which is currently comprised of only 
14 sources. NMBS-II, a shallow-wide accepted NOAO Survey Program specifically 
designed to further constrain the population of very massive galaxies at 
$z>2$, will image an area of sky a factor of $\sim10$ larger than NMBS, 
resulting in a significant increase in the number of very massive galaxies 
out to $z\sim3.5$ with accurate photometric redshifts and well-sampled SEDs.

Follow-up multi-object spectroscopic observations in both the optical and in 
the NIR are of vital importance to confirm the redshifts and to better 
characterize the properties of the most massive galaxies at $z=3.5$, 
including AGN and Lyman-$\alpha$ emitter fractions, AGN and/or starburst 
contamination of the optical-to-MIR SEDs, superwind outflows, star formation 
rates, and mass-to-light ratios. However, to probe the rest-frame wavelength 
regime red-ward of $\sim5000$~\AA, and hence to robustly constrain the star 
formation histories and SFRs, measure metallicities, absorption lines and 
velocity dispersions from rest-frame optical features, will require NIRSPEC 
on the {\it James Webb Space Telescope}. The estimated total IR luminosities 
typical of HLIRGs make the very massive galaxies at $3 \leq z < 4.0$ in our 
sample ideal candidates for follow-up observations with the 
{\it Atacama Large Millimeter Array} (ALMA). ALMA will be crucial in 
constraining the amount of dust and gas in these systems, as well as 
discriminating between dust-enshrouded star-formation and obscured AGN 
activity. Moreover, it will also allow for measurements of the kinematics 
in these systems, providing an independent estimate of the dynamical masses 
of the most massive galaxies at $3.0 \leq z < 4.0$.

Finally, to fully characterize the population of galaxies at $3<z<4$, the 
analysis performed in this work has to be extended to lower stellar masses. 
This will necessarily require very deep imaging with NIR medium band-width 
filters to provide very accurate photometric redshifts and well-sampled SEDs 
down to faint $K$-band magnitudes.


\acknowledgments

We are grateful to the anonymous referee whose comments and suggestions 
helped improving significantly this paper. Ron Probst and the NEWFIRM team 
are thanked for their work on the instrument and help during the 
observations. This paper is partly based on observations obtained with 
MegaPrime/MegaCam, a joint project of CFHT and CEA/DAPNIA, at 
the Canada-France-Hawaii Telescope (CFHT) which is operated by the National 
Research Council (NRC) of Canada, the Institut National des Sciencie de 
l'Univers of the Centre National de la Recherche Scientifique (CNRS) of 
France, and the University of Hawaii. This work is based in part on data 
products produced at TERAPIX and the Canadian Astronomy Data Centre as part 
of the CFHT Legacy Survey, a collaborative project of NRC and CNRS. Support 
from NSF grants AST-0449678 and AST-0807974, and NASA LTSA NNG04GE12G is 
gratefully acknowledged.


\end{document}